\documentclass[aps,twocolumn,showlabels,showrefs,amsmath,amssymb,prl,superscriptaddress,floatfix,colors]{revtex4-1}

\usepackage{graphicx}
\usepackage{dcolumn}
\usepackage{bm}
\usepackage{amssymb}
\usepackage{hyperref}
\usepackage{multirow}
\usepackage{color}
\usepackage[normalem]{ulem}

\begin{document}

\title{Motility-Induced Microphase and Macrophase Separation in a Two-Dimensional Active Brownian Particle System}


\author{Claudio B. Caporusso}
\affiliation{Dipartimento  di  Fisica,  Universit\`a  degli  Studi  di  Bari  and  INFN,
Sezione  di  Bari,  via  Amendola  173,  Bari,  I-70126,  Italy}

\author{Pasquale Digregorio}
\affiliation{Dipartimento  di  Fisica,  Universit\`a  degli  Studi  di  Bari  and  INFN,
Sezione  di  Bari,  via  Amendola  173,  Bari,  I-70126,  Italy}
\affiliation{CECAM  Centre  Europ\'een  de  Calcul  Atomique  et  Mol\'eculaire,
Ecole  Polytechnique  F\'ed\'erale  de  Lausanne,  Batochimie,  Avenue  Forel  2,  1015  Lausanne,  Switzerland}

\author{Demian Levis} 
\affiliation{Departement de Fisica de la Materia Condensada, Facultat de Fisica, Universitat de Barcelona, Mart\'{\i}  i  Franqu\`es  1,  E08028  Barcelona,  Spain}
\affiliation{UBICS  University  of  Barcelona  Institute  of  Complex  Systems,  Mart\'{\i}  i  Franqu\`es  1,  E08028  Barcelona,  Spain}


\author{Leticia F. Cugliandolo}
\affiliation{Sorbonne Universit\'e, Laboratoire de Physique Th\'eorique et Hautes Energies, CNRS UMR 7589, 
4 Place Jussieu, 75252 Paris Cedex 05, France}
\affiliation{Institut Universitaire de France, 1 rue Descartes, 75005 Paris France}
 
\author{Giuseppe Gonnella}
\affiliation{Dipartimento  di  Fisica,  Universit\`a  degli  Studi  di  Bari  and  INFN,
Sezione  di  Bari,  via  Amendola  173,  Bari,  I-70126,  Italy}

\email[email: ]{name@}


\begin{abstract}
As a result of non-equilibrium forces, purely repulsive self-propelled particles undergo macro-phase separation between a dense and a dilute phase. We present a thorough study of the ordering kinetics of such Motility-Induced Phase Separation (MIPS) in Active Brownian Particles in two-dimensions, and we show that it is generically accompanied by micro-phase separation. 
The growth of the dense phase follows a law akin to the one of liquid-gas phase separation.
However, it is made of a mosaic of hexatic micro-domains whose size does not coarsen indefinitely,  leaving behind a network of extended topological defects from which microscopic dilute bubbles arise.  The characteristic length of these finite-size structures 
increases with activity, independently of the choice of initial conditions. 
\end{abstract}

\maketitle

Active 
systems are ubiquitous in Nature. Driven out of equilibrium
by the consumption of energy from the environment,
they cannot be described using the tools of equilibrium statistical physics,
and present intriguing collective behavior~\cite{MarchettiRev,WinklerRev}. One such peculiarity is that, 
at sufficiently high activity,
 their constituents cluster in the absence of attractive interactions. 
A steady state with a dense droplet immersed in 
a dilute background can thus be reached in systems of purely-repulsive spherical particles by Motility-Induced Phase Separation (MIPS)~\cite{CatesRev}. 

Arguably, the simplest microscopic active matter model is one of
self-propelled spheres undergoing rotational 
diffusion, and excluded volume interactions only. This is the Active Brownian Particles (ABP) model that 
exhibits a very rich phase diagram, including MIPS, especially in two dimensions ($2d$)~\cite{Romanczuk2012, Bialke2012cryst, Fily2012, Stenhammar2014,  Joan, Speck15, Redner13, PRLino, KKK,defectsLino, PaliwalDijkstra, BennoLowen, CapriniVelocities}. 
In particular, the special role played by hexatic order in these systems was discussed in~\cite{Redner13,PRLino,KKK,defectsLino,PaliwalDijkstra},  
and the existence of a phase with such order was exhibited~\cite{PRLino}. In parallel to particle based models, the 
large scale and long time behavior of systems exhibiting MIPS  {was} addressed with adaptations of the 
Cahn-Hilliard approach~\cite{Stenhammar13,wittkowski2014,Speck14,Speck15,Tjhung18}. 

While the non-equilibrium phase diagram of (hard) ABP is well established~\cite{PRLino}, 
the dynamics across the various phase transitions, how is 
the dense droplet formed via  MIPS, and which is its actual nature, still need clarification. 
A relevant question to ask is 
whether the hexatic ordering helps or interferes
with the simple particle aggregation. Moreover, whether 
the droplet behaves as a $2d$  {hexatic, liquid or else}, featuring or not gas bubbles inside~\cite{Tjhung18},  
are important issues that have to be elucidated. 

\begin{figure}[b!]
\centering 
\hspace{1.5cm}
\includegraphics[width=8.cm]{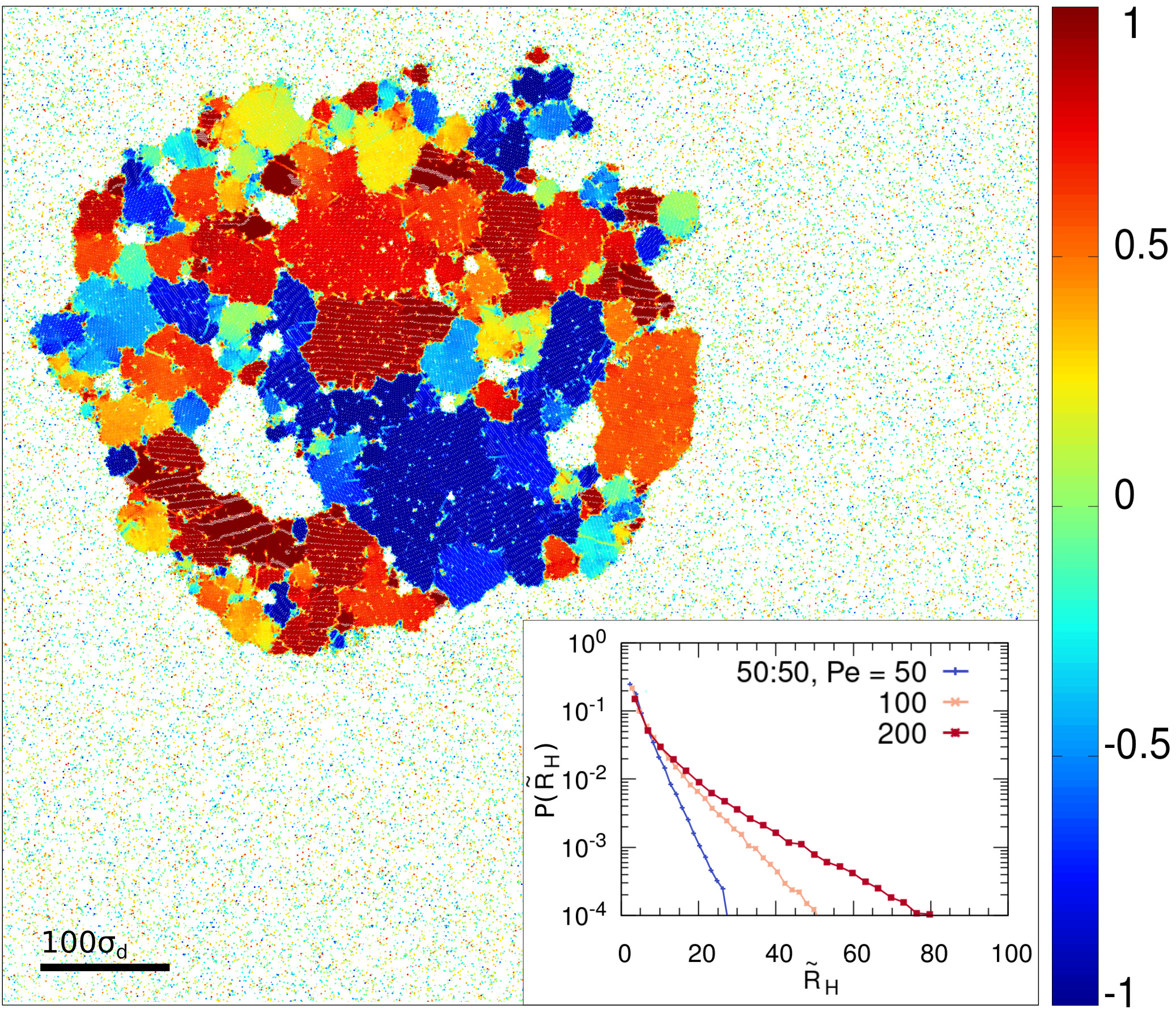} 
\vspace{0.25cm}
\caption{{\bf 
Hexatic domains and bubbles  in a 
MIPS droplet.}
A steady-state snapshot of $512^2$ ABP at moderate density ($\phi=0.25$) and high activity (Pe = 200) showing 
a \emph{macro}-droplet made of a mosaic of hexatic \emph{micro}-domains,  delimited by clusters of topological defects 
from which \emph{micro}-bubbles arise. 
{Colors indicate the projection of the particles' hexatic parameter $\psi_{6,j}$ 
(see the definition in the text) onto the direction of its global average $\Psi_6=\frac{1}{N} \sum_j\psi_{6,j}$.}
Insert: stationary distribution of hexatic domain radii  {$\tilde R_H$} 
for three Pe and equal fraction of dense and dilute 
coexisting phases. 
}
\label{fig:snap0}
\end{figure}

In this Letter we address these points and in so doing we clarify 
the origin of the cavitation gas bubbles 
recently predicted with a continuum description~\cite{Tjhung18}. Using extensive numerical simulations (more than $10^6$ 
ABP)  we 
  {exhibit} and characterize several dynamic regimes: multi-nucleation, condensation and aggregation, and 
 coarsening fulfilling dynamic scaling,  see movie~1 in the SM~\cite{SM}.
Furthermore, we show that, asymptotically, the \emph{macro}-droplet self-organizes into a mosaic 
of hexatic \emph{micro}-domains, see Fig.~\ref{fig:snap0},  {differently from what happens in equilibrium co-existence.} 
The hexatic domains do not coarsen to reach the droplet 
size but rather saturate to a microscopic  {though relatively large scale which}
can be directly controlled by self-propulsion.  
Different hexatic domains are delimited by clusters of topological defects, leaving behind regions of lower density   from which {\it micro}-\-bubbles  {pop up}.  
Devising generic mechanisms to control the spatiotemporal organization of active matter into structures that do not coarsen constitutes a central challenge of current research  {which}  has been tackled with self-propelled particles with chemotactic  \cite{BennoCluster, BennoCluster2}, competing  \cite{ChantalFrenkel,PacoChantal}, or anisotropic (polar or nematic)  \cite{SolonChate, BennoPRL, GiomiPierce} interactions, among others. Here we  {exhibit} 
a structure   that does not coarsen, associated to the hexatic order,  {in the simplest active particle model}.

We consider  $N$  particles at positions ${\bold{r}}_i$ 
in an $L\times L$ box with periodic boundary conditions evolving {\it via}
\begin{equation}
\begin{array}{rcl}
	\label{eq:langevin}
	\gamma\dot{\bold{r}}_i &=& F_{\rm act} \bold{n}_i- \sum_{j(\neq i)}{\boldsymbol{\nabla}}_iU(r_{ij}) + \sqrt{2 \gamma k_B T}\bm{\xi}_i \; ,
	\\
	 \dot{\theta}_i &=& \sqrt{2 D_{\theta}}\eta_i 
	 \,,
	 \end{array} 
\end{equation}
(see \cite{PRLino} for details)
where $F_{\rm act}$ is the self-propulsion force acting along $\bold{n}_i=( \cos{\theta_i(t)},\sin{\theta_i(t)})$, and 
 $U(r)=4\varepsilon [({\sigma}/{r})^{64}-({\sigma}/{r})^{32}]+\varepsilon$ if $r< \sigma_d=2^{1/32}\sigma$ and $0$ otherwise, with $r_{ij}=|{\bold{r}}_i-{\bold{r}}_j|$. 
The  components of $\bm{\xi}$ and $\eta$  are zero-mean  and unit variance independent 
white Gaussian noises.
The units of time, length and energy  are $\tau=D^{-1}_\theta$, $ \sigma_d$ and $\varepsilon$, respectively.  
We fix $\gamma=10$, $k_BT=0.05$ and $D_\theta = 3 k_BT/(\sigma^2_d \gamma)=0.015$. 
We perform quenches by suddenly turning on activity, quantified by the P\'eclet number Pe = $F_{\rm act} {\sigma_d}/(k_BT)$,
deep in the MIPS coexistence region~\footnote{Here, the MIPS critical point is located at (Pe, $\phi)\approx(32,0.6)$}, on a homogeneous configuration at a given packing fraction $\phi =\pi{\sigma^2_d}N/(4L^2)$  {and other synthetic initial states.}
We monitor the 
characteristic length scales of phase separation, hexatic ordering and cavitation bubbles
 {to build a full picture of the ordering process.}

The dense phase is formed through a rather complex process
 {that we review and complete to set the stage for our study.}
\begin{figure}[h!]
\centering 
\includegraphics[width=8.6cm]{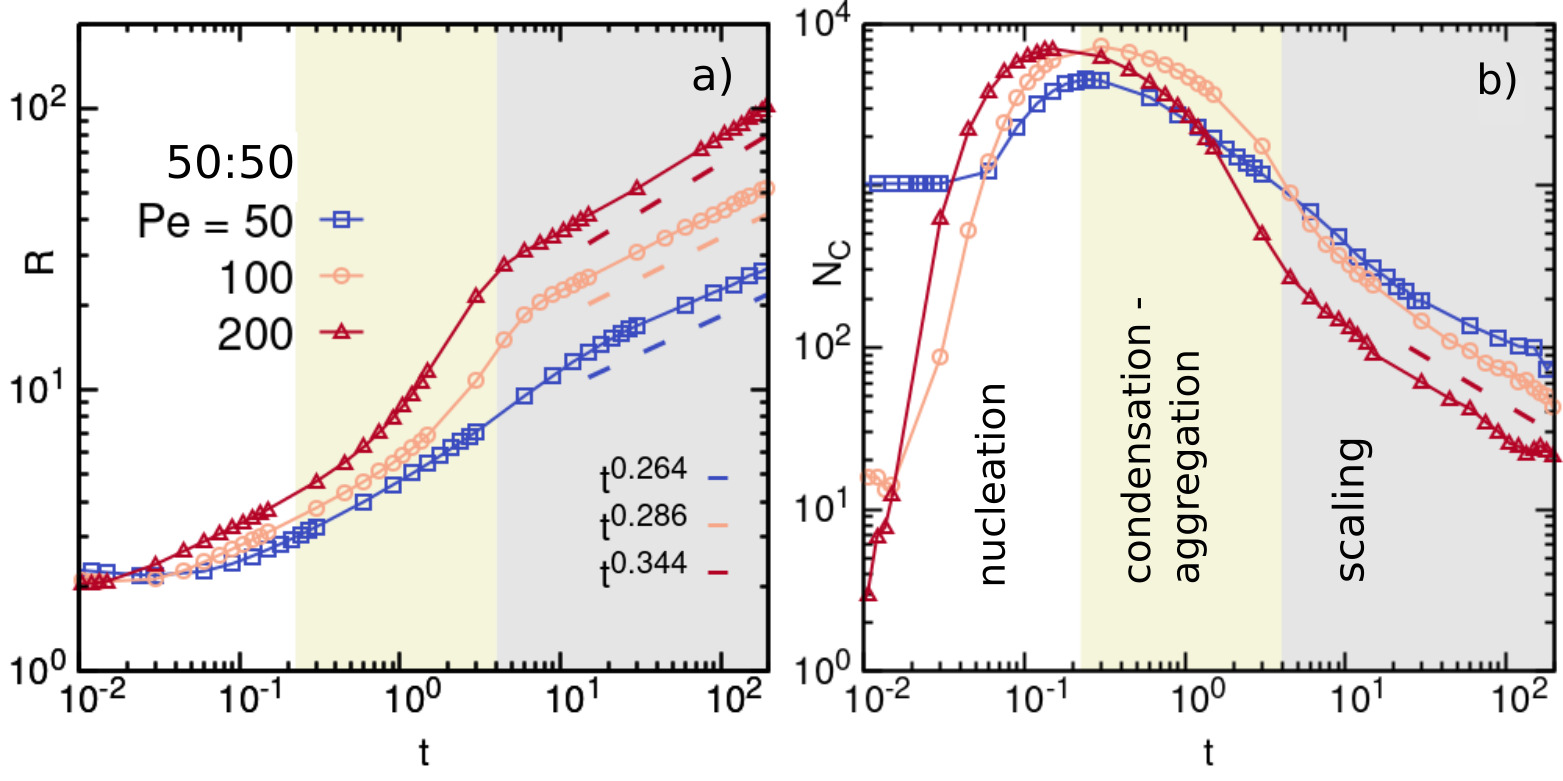}
\vspace{-0.25cm}
\caption{
{\bf Dense phase formation.}
(a) Typical {dense phase} size from $S(k,t)$  
and (b) number of clusters, both as a function of time, after a quench into the MIPS coexistence region, in a system with $N=1024^2$ particles
at different  Pe on 
the curve with equal surface fraction of dense and
dilute phases. 
Three time regimes are discerned: nucleation, condensation and aggregation, and scaling. 
}
\label{fig:growing-length}
\end{figure}
 {The first moment of the structure factor, $R(t)=\pi/ [\int d k \, k \, S(k,t)/\int d k \, S(k,t) ]$, 
with $S(\bm k,t) = N^{-1} \sum_{ij} e^{{\rm i} {\bm k} ({\bf r}_i - {\bf r}_j)}$, serves to estimate its  length scale and 
is plotted in Fig.~\ref{fig:growing-length}(a) for three Pe  (lying 
on the 50:50-curve defined as the set of points  for which the system de-mixes into equal 
dense and dilute portions). First, there is multi-nucleation of tiny droplets
and $R$ is small and roughly independent of Pe. A crossover to a regime in which small droplets evaporate while larger 
ones grow by condensation and aggregation is favored by increasing Pe.
A scaling regime, similar to Ostwald ripening, establishes next and $R$ is algebraic with a universal exponent $1/z$ that is closer to the 
expected $z=3$ \footnote{Even in kinetic Ising models for which 
$z=3$ is proven, it is notably difficult to measure 
it numerically~\cite{Tartaglia18}.} of phase separation~\cite{BrayRev} (or diffusion limited regular cluster aggregation~\cite{Leyvraz03}) than the values estimated in previous works~\cite{Redner13,Stenhammar2014,BennoLowen}.
The dynamic scaling hypothesis~\cite{BrayRev} is fulfilled in this regime~\cite{SM}.
 Finally, the size of the dense cluster
saturates to a value that grows with the system size $L$. 
The total number of particle clusters, $N_C$ in Fig.~\ref{fig:growing-length}(b)
(definition in~\cite{SM}), illustrates the nature of the first two regimes. During nucleation $N_C$ 
grows fast until a maximum,  which signals the crossover towards
aggregation and its progressive fast decay. In the  scaling regime a fit to 
the Pe = 200 data yields $ N_C\sim t^{-0.59}$, consistently with the growth of $R$.
}

\begin{figure}[h!]
\centering 
\includegraphics[width=8.6cm]{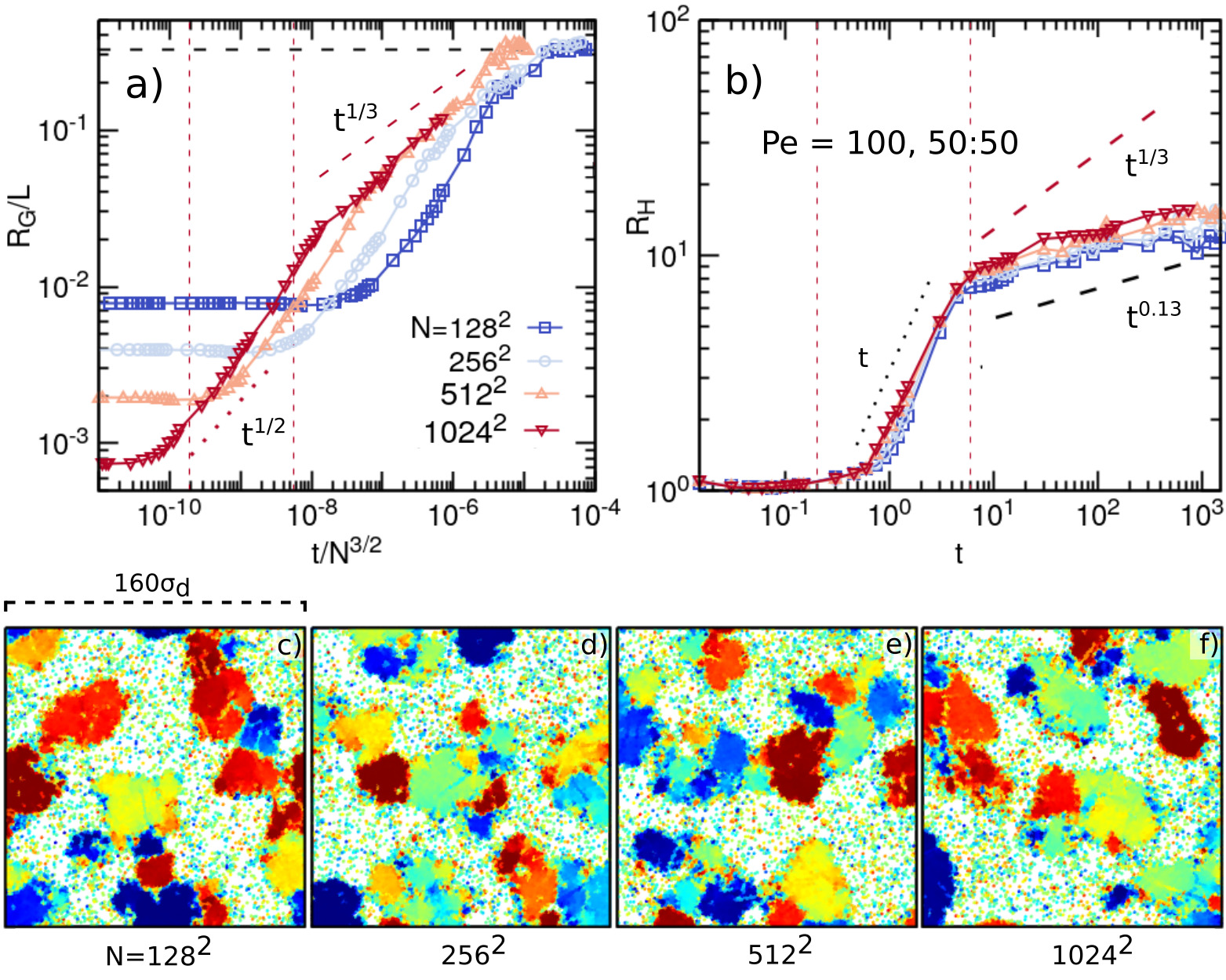}
\vspace{-0.5cm}
\caption{
{\bf Macro-droplet against hexatic micro-domains.}
Growth of the dense phase radius of gyration, $R_G$, normalized by the system size $L$ 
(a) and the hexatic length $R_H$  (b)  for Pe = 100 and different $N$ on the 50:50-curve.
In (a) $R_G/L$ saturates  at $\sqrt{5/48}$ 
(a droplet that occupies half the available surface).
In (b) we represent the  slow growth $t^{0.13}$, 
and  $t^{1/3}$ for comparison. 
The vertical dotted lines mark the crossover between the three regimes in the largest system.
Below, a zoom over a sector of these
systems with linear size $160\, \sigma_d$ taken at $t=15$ (scaling regime) is shown.}
\label{fig:Ndep}
\end{figure}

\begin{figure}[b!]
\centering 
\includegraphics[width=8.6cm]{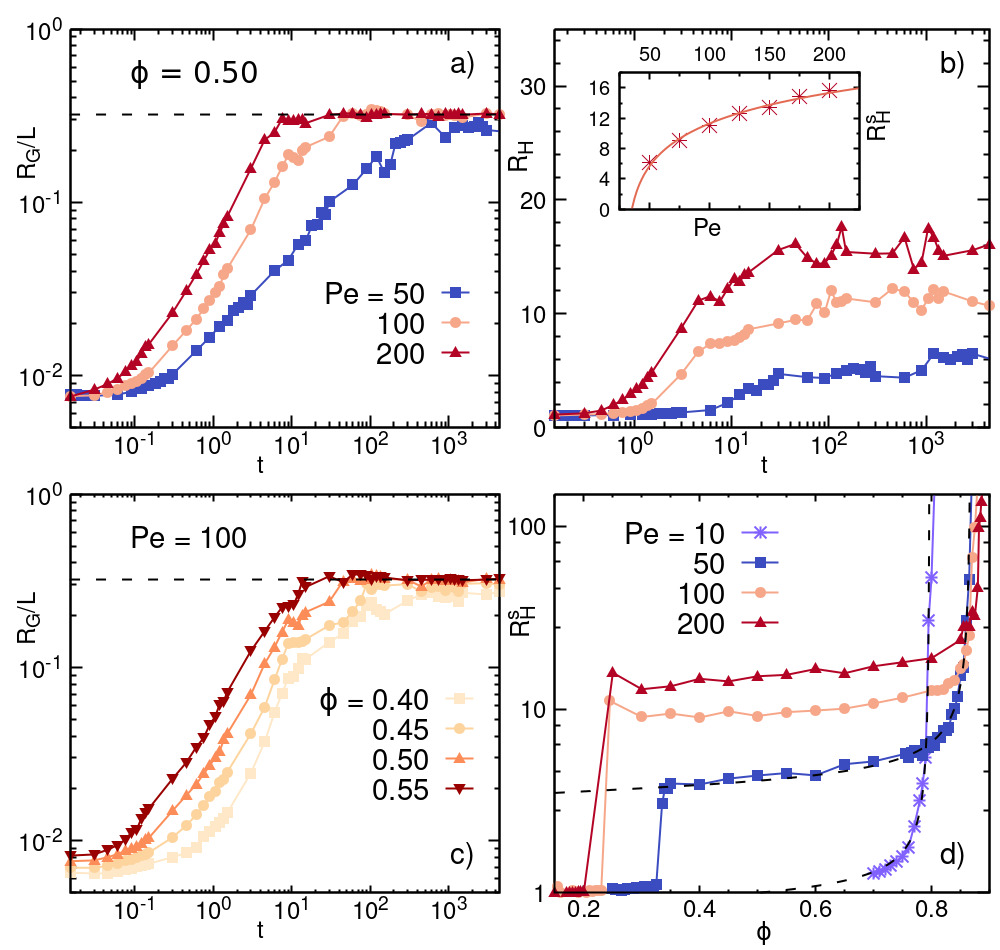} 
\caption{{\bf Parameter dependencies of the dense phase and hexatic growing lengths.}
$N=128^2$ at fixed $\phi=0.5$ and varying Pe (a),\,(b); at fixed Pe = 100 and varying $\phi$ (c)
 {and the asymptotic $R_H^s$ for Pe = 50, 100, 200 (in MIPS) and Pe = 10 (before MIPS)
as a function of $\phi$ (d).  An exponential growth $a \exp[b\,(\phi^{-1}-\tilde{\phi}^{-1})^{-1/2}]$ is shown in dotted lines 
(for Pe = 10: $a= 0.778$, $b=0.214$, and $\tilde{\phi}=0.797$; for Pe = 50: $a=3.172$, $b=0.232$, and $\tilde{\phi}=0.868$).} 
Inset in (b): dependence of the asymptotic value $R_H^s$ on Pe; the continuous line is 
$R_H^s\propto\ln[\text{Pe}-a]+b$ with $a=29.01$ and $b=-1.79$ (vanishing at the transition point). 
 }
\label{fig:Pephidep}
\end{figure}

Another measure of the dense phase size is provided by the average radius of gyration of the clusters, 
$R_G^2 = \langle N^{-1}_d  \sum_{i\in d} ({\mathbf r}_i - {\mathbf r}_d^{\rm cm})^2\rangle$, 
with ${\mathbf r}_i $ the position of the $i$th particle among the $N_d$ ones in the cluster, and 
${\mathbf r}_d^{\rm cm}$ the position of the cluster center of mass. $\langle \dots \rangle$ is an average over all clusters.
Its behavior, in Fig.~\ref{fig:Ndep}(a), is analogous  to the one of $R$.
The dotted vertical lines 
locate the crossovers between the  time-regimes in Fig.~\ref{fig:growing-length}, with
a slightly more extended intermediate one in which
$R_G$ raises with the $t^{1/2}$ of Brownian aggregation~\cite{Leyvraz03} to later accelerate pushed by activity, and then
crossover to $t^{1/3}$ when the mass of the gas reaches its constant target value~\cite{SM}.
$R_G$  saturates to 
$R_G^s\propto L$ after the relaxation time  $N^{z/2}$. 
For the largest  {system}, $N=1024^2$,  {$z$ is} very close to 
 {$z=3$}. 
At fixed $N$ and $t$, $R_G$  increases monotonically with $\phi$ and Pe, 
since the fraction of the system in the dense phase also does, see Fig.~\ref{fig:Pephidep}(a),\,(c).  

From the previous study, depending on the quench, one may expect any kind of length scale associated to a relevant order 
parameter to either grow to reach a macroscopic value $\propto L$ or, on the contrary, relax to a vanishing one.  
We now  show that  
this does not apply to the hexatic order of ABP in the MIPS regime.
 {Moreover, we demonstrate that several so-far ignored 
features are different from the ones in equilibrium co-existence.}

We attach a hexatic order parameter to each particle, $\psi_{6j}=N_j^{-1}\sum_{k\in\partial_j}e^{i6\theta_{jk}}$, 
with  $\theta_{jk}$  the angle formed by the segment that connects the center of the $j$th disk and the one of its $k$th,  
out of $N_j$, nearest neighbors found with a Voronoi construction. 
The colors in the snapshots in Figs.~\ref{fig:snap0}  {\&} \ref{fig:Ndep} represent different 
{local}
hexatic order, {defined as the projection on their global average, a continuous variable in $[-1,1]$.}
The clusters that aggregate do not necessarily share the same $\psi_{6j}$, and grain boundaries progressively 
appear in the growing dense phase. Whether they heal or not is the issue that we now address.

We identify the hexatic domains according to the argument of $\psi_{6j}$  
or by the gradient of its modulus, coarse-grained over a small cell, 
and we find equivalent results with the two methods~\cite{SM}.
The average (over domains) gyration radius, $R_H$,  is exposed in Fig.~\ref{fig:Ndep}(b)  for different $N$,  
and in Fig.~\ref{fig:Pephidep}(b),\,(d) for different Pe.
$R_H$ is quite independent of the system size. In the intermediate regime
$R_H \simeq t$. At a sharp crossover  {concomitant with the entrance in the scaling regime}
the evolution slows down remarkably to $R_H\simeq t^{0.13}$, a law 
confirmed by the analysis of the corresponding structure factor {~\cite{SM}}. Similar small exponents 
were found in the growth of order in certain pattern formation processes~\cite{Cross93}
and in the growth of hexatic order in block co-polymer systems~\cite{Vega05}.
Later $R_H$ approaches a finite limit, $R_H^s$, proving the \emph{arrested coarsening} 
of hexatic order, see Fig.~\ref{fig:Ndep}(b).  
The snapshots in Fig.~\ref{fig:Ndep}(c)-(f) show that hexatic domains have roughly the same size in systems of different dimension. 
The asymptotic $R_H^s$ increases with Pe at fixed $\phi$ but does not 
change appreciably with $\phi$ at fixed Pe, Fig.~\ref{fig:Pephidep}(d), meaning that self-propulsion controls the size of the hexatic 
micro-domains.  {This is confirmed by  the exponential distributions of the individual  {$\tilde R_H$} 
 in  Fig.~\ref{fig:snap0} with average compatible with the data in the inset in Fig.~\ref{fig:Pephidep}(b).}
 
 The dense component in MIPS is not like the hexatic component in 
 equilibrium co-existence. Indeed, the map of local hexatic order  at Pe = 0 is not of the multicolor 
 kind in Fig.~1 but it has just one reddish denser component, the single hexatic domain,
 see Fig.~S12 in the SM. This reflects the fact that the hexatic co-existing phase in equilibrium is inherited from a proper 
 hexatic phase with a diverging correlation length. Moreover,
{our} equilibrium or active liquid is also homogeneous from the hexatic order point of view, with a very low $R_H$. 
 This is illustrated by the Pe = 10 data in Fig.~\ref{fig:Pephidep}(d), which show $R_H \sim 1$ until very close to the 
transition where an exponential divergence  \`a la BKT takes over  {in this case}.  
For the three sets of data-points within MIPS, the mosaic length 
{$R^s_H$} is almost constant and much larger than the one of a homogeneous liquid,
 until an exponential takes over close to the upper border of MIPS.
{One may wonder how this compares to hexatic  ordering in the phase separation of equilibrium fluids of attractive particles. 
 This is an issue that has only recently been addressed, with the observation that 
 attractive interactions generically destabilize hexatic order~\cite{Kim19,Pica20}; 
 therefore,  a stationary mosaic structure (with bubbles, see below) as the one in Fig.~\ref{fig:snap0} is not expected in 
 these systems.}

Finally,  we did not find any correlation between the local hexatic and velocity fields nor a finite velocity correlation length, as recently  {reported}
in a similar (though athermal) ABP model  as $\tau={D_\theta^{-1}}$ is increased at fixed $v=F_{\rm act}/\gamma$~\cite{CapriniVelocities}.

\begin{figure}[t!]
\vspace{-0.25cm}
\centering 
\includegraphics[width=8.6cm]{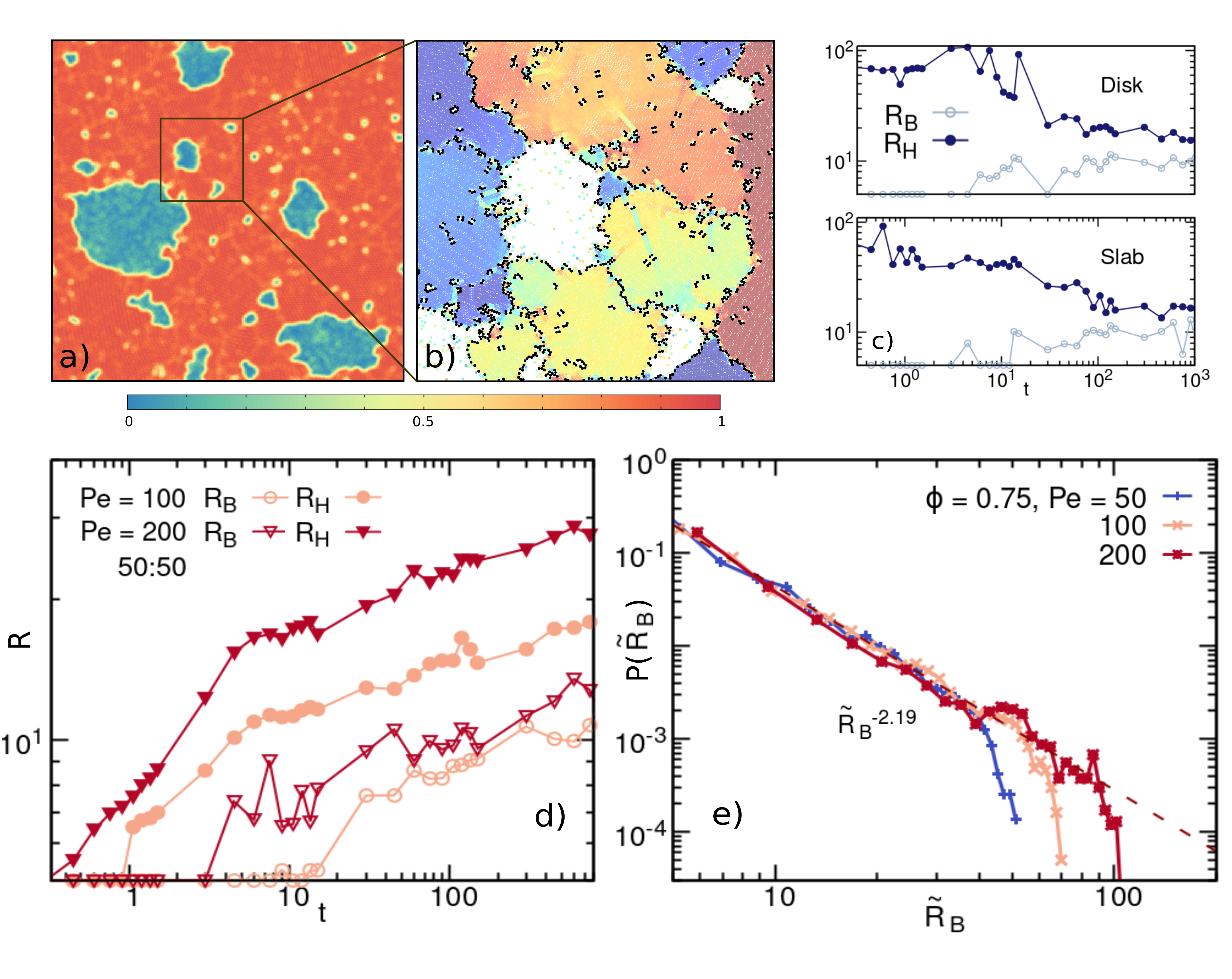} 
\vspace{-0.6cm}
\caption{{\bf Gas micro-bubbles.}
 {Snapshot at $\phi=0.75$,  Pe = 200 (a). The  color scale (below) is associated to the local density. 
In (b) a zoom over a sector of  (a)} showing hexatic domains 
delimited by clusters of defects (represented by black dots), where bubbles (dilute regions) 
are likely to emerge.  {(c)} 
Evolution of $R_B$ and $R_H$ from an initial hexatically ordered disk or slab  at Pe = 100 and $\phi=0.5$  (see movies 2 and 3 \cite{SM}).
(d) Growth of the bubble typical radius $R_B$ (empty symbols) confronted to the one of the hexatic order $R_H$ (filled symbols) for two Pe. 
(e) Steady-state distribution of bubble radii  {$\tilde R_B$} (parameters given in the key).
}
\label{fig:bubble}
\end{figure}

Besides the emergence of hexatic micro-domains, 
MIPS is accompanied by the formation of gas bubbles. 
As illustrated in Fig.~\ref{fig:bubble}(a)-(b)
and even more clearly in movies 1-3~\cite{SM}, all bubbles have the same density (set by the lower branch of the  {MIPS binodal, see Fig.~S3~\cite{SM})}. 
Their averaged size, measured, for example, from their radius of gyration, $R_B$, increases with Pe
(see~\cite{SM} for details). 
Just as the hexatic micro-domains,  bubbles do not coarsen indefinitely. Figure~\ref{fig:bubble}(d) shows that
$R_B$ is delayed with respect to $R_H$, but after the transient the two quantities grow parallelly 
in the log-log representation indicating that 
they follow a similar trend until eventual saturation to a Pe-dependent value~\cite{SM}.  
The steady state-distribution, Fig.~\ref{fig:bubble}(e), decays algebraically,  {$P(\tilde R_B) \sim {\tilde R}_B^{-2.19}$} for $\phi=0.75$, 
independently of Pe, until a Pe-increasing cut-off   {${\tilde R}_B^*({\mbox{Pe}})$}. 
Interestingly, the system approaches the same finite $R^s_H$ and $R^s_B$ independently of the initial condition.
This fact can be visualized in movies 2 \& 3~\cite{SM} where a disk and a slab drop with uniform hexatic order are used 
as initial states, respectively. In both cases the dense component progressively breaks into finite size domains 
with different orientation, leaving space for bubbles at their interfaces. 
Consistently,  Fig.~\ref{fig:bubble}  {(c) shows} the rise of $R_B$,  and corresponding drop-off of $R_H$, 
which approach asymptotic values that are consistent with the ones obtained after a quench from disordered initial conditions. 
 
Micro-bubbles have been predicted by a continuum model of MIPS~\cite{Tjhung18},  
but their existence in particle-systems has not been studied yet.
 Micro-phase-sep\-a\-ra\-tion in the continuum model takes place in a specific parameter regime.
However, such parameters (in particular the one quantifying the term leading to the micro-bubbles) 
cannot be readily translated in terms of the $\phi$ and Pe of ABP. 
 Here, contrarily to the mean-field calculations in~\cite{Tjhung18}, we observe micro-bubbles as long 
 as MIPS takes place, both at moderate and high densities, from Pe = 40 to Pe = 200 
  (see Figs.~\ref{fig:snap0} \& \ref{fig:bubble} at $\phi=0.25$ and $0.75$, respectively, and \cite{SM}), 
  and show that their size can be controlled by Pe.  {In addition, our size distribution is algebraic 
  while the one in~\cite{Tjhung18} is peaked at a favored 
  length scale.}
  
The microscopic origin of micro-bubbles in ABP can be
tracked down to the presence of topological defects (mis-coordinated particles with more or less than the 6 neighbors)
mostly localized at the boundaries of the hexatic micro-domains, see  {Fig.~\ref{fig:bubble}(b)} and~\cite{defectsLino}.
A fluctuation is thus more likely to generate a bubble at a grain boundary than within a hexatic domain.
Some of the bubbles thus generated 
quickly decay, while others grow and have very long life-times ($\sim 10-100\,\tau$), of the same order as the 
reorganization time scale of the hexatic domains.   These features are displayed in the movies~\cite{SM}.
The emergence of a finite Pe-dependent length-scale associated with the hexatic order must be responsible for the presence of 
micro-bubbles as suggested by the growth kinetics of $R_H$ and $R_B$, which 
evolve at the same very slow rate. Furthermore, the cut-off algebraic distribution of bubble radii, Fig.~\ref{fig:bubble}(e), 
is intimately related to the same kind of statistics found for topological defect clusters~\cite{defectsLino}.

Summarizing, we monitored the Motil\-i\-ty-In\-duced Phase Separation (MIPS) of large systems of ABP.
On top of the dense-dilute phase separation, governed by a $t^{1/3}$ growing length in the scaling regime,
$2d$ MIPS involves another ordering mechanism controlled  by activity,  
giving rise to a new finite characteristic length associated to hexatic order.  As a result,  
the dense phase breaks into a mosaic of differently oriented patches.
The analysis of hexatic order thus provides a new means of controlling the self-organization of active particles. 
The ability of active systems to sustain non-equilibrium structures and control coarsening has been reported in a variety of cases 
(Janus colloids, bacteria, motility assays, etc.) \cite{Cecile2012,GiomiPierce, Schaller2010,JulicherRev2019} but has remained elusive in simple models of spherical self-propelled particles. We showed that self-propulsion and excluded volume are sufficient to arrest 
orientational coarsening and stabilize finite-size structures. 
 {This is an essential out-of-equilibrium effect driven by activity.
In addition,} we proved that gas bubbles naturally appear 
in the interstices between different hexatically ordered patches, where topological defects favor their emergence. 
The bubbles' growth follows the same (delayed) pace as the one of the hexatic patches 
and their size also saturates to a finite value that  {increases with} Pe but  {does not significantly depend on} 
$\phi$. 
The statistics of cavitation bubbles is in close relation to the one of clusters of topological defects.
It would be interesting to associate these features to measurements of local pressure but these are notably difficult to 
carry out and lie beyond the scope of this study.

\vspace{0.25cm}

\noindent
{\it Acknowledgments.}
We acknowledge access to the MareNostrum Supercomputer at the BSC,  Lenovo NeXtScale MARCONI at 
CINECA (Project  INF16-field\-turb)  under  CINECA-INFN  agreement
This research is  supported by MIUR project PRIN 2017/WZFTZP ``Stochastic forecasting in complex systems".
D. L. acknowledges funding from JIN project RTI2018-099032-J-I00  (MCI/AEI/FEDER, UE). 


\pagebreak
\widetext
\begin{center}
\textbf{\large Supplementary Material - Motility-Induced Microphase and Macrophase Separation in a Two-Dimensional Active Brownian Particle System
}

\end{center}

\setcounter{equation}{0}
\setcounter{figure}{0}
\setcounter{table}{0}
\setcounter{page}{1}

\renewcommand{\theequation}{S\arabic{equation}}
\renewcommand{\thesection}{S\arabic{section}}
\renewcommand{\thefigure}{S\arabic{figure}}

\vspace{2cm}

In this Supplemental Material we show additional information on the MIPS process undergone by Active Brownian Particles 
(ABP), following the dynamics defined in Eq.~(1) in the main text, together with a detailed description of the 
numerical methods used for the analysis presented  in the main text. The document is organized as follows: Section I 
presents three videos that illustrate the dynamic mechanisms at work. In Sec. II we provide snapshots, drawn in the form of heat maps of the local density,  to further illustrate the structure of the system in the Motility Induced Phase Separation  (MIPS) regime, and make clear the existence of cavitation bubbles. In Sec. III we show how the phase ordering kinetics of MIPS fulfills the dynamical scaling hypothesis, and prove that the structure factor exhibits the expected 
small wave-vector dependence and Porod's law beyond the first peak.
In Sec. IV we describe the clustering algorithms that we used to identify the different characteristic length scales discussed in the Letter, based on the analysis of the local surface fraction and the local hexatic order parameter. 
 {Section V discusses the hexatic order growth and especially its dynamic scaling 
properties and, finally, Sec. VI presents an analysis of the hexatic component in passive co-existence.}

\vspace{0.5cm}

\section{I. Videos}
\label{sec:videos}

\begin{figure}[h!]
\vspace{0.75cm}
\includegraphics[width=\textwidth]{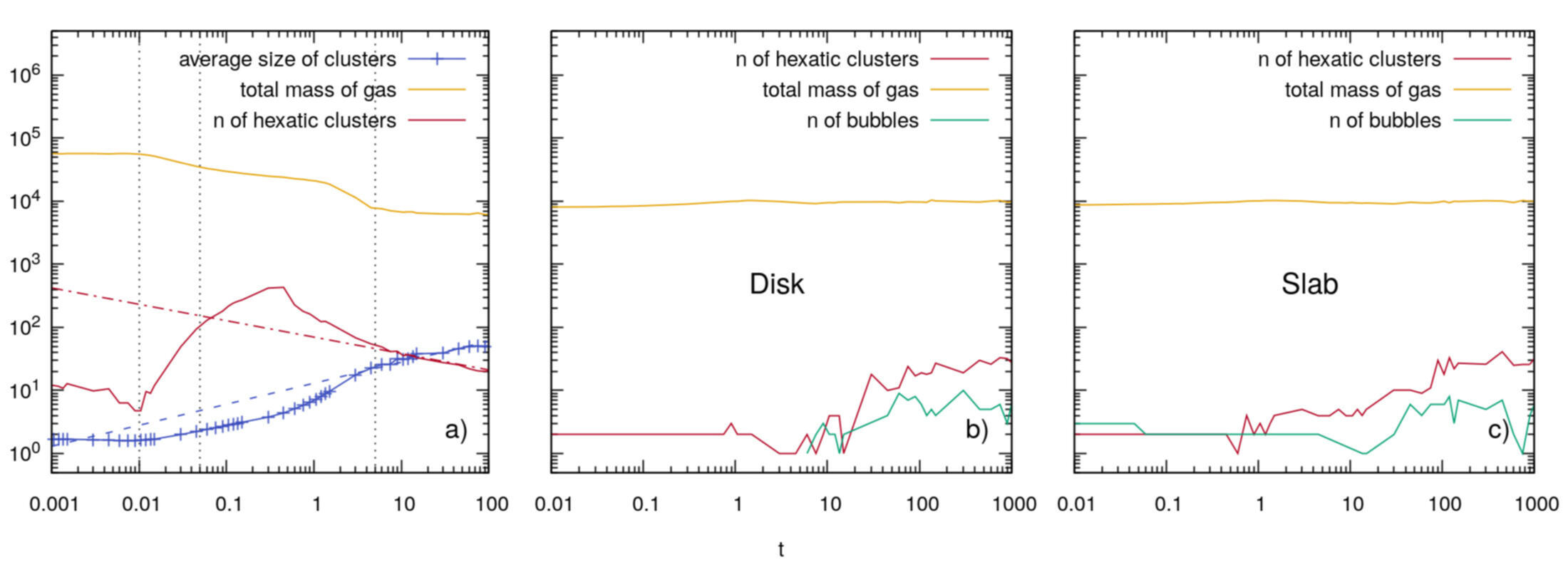}
\caption{(a) Average radius of droplets (blue symbols) as a function of time, 
evaluated as described in Sec III, with approximate power-law growth $R(t)\sim t^{1/3}$ in the scaling regime (dotted blue line). The yellow curve represents 
the total number of particles $M_{gas}$ belonging to the gaseous phase, which, notably, reaches a plateau in the scaling regime. The red line is the evolution in time of the total number of hexatic clusters. The dashed red line represents the power-law decaying of the latter within the scaling regime, as estimated from the absence of particle exchange between the two phases and $n_H \sim M_{dense}/R_H^2 \sim t^{-0.26}$, given the measure of the growing exponent for the average hexatic radius shown in the main text. Data refer to the dynamics shown in Movie1, with a random homogeneous initial condition. Finally, the dotted vertical lines delimit the 
three regimes: nucleation, condensation and aggregation, and scaling.
(b) Number of hexatic clusters (red line) and bubbles (green line), and total mass of the gaseous phase (yellow line), corresponding to the dynamics of Movie2, with an hexatically ordered disk as initial condition. (c)  Same quantities for the dynamics of Movie3, with a slab with hexatic order as initial condition.}
\label{fig:figure_video}
\end{figure}

Three videos reproduce the system dynamics starting from different initial states. Some relevant quantities concerning the dynamics of the three videos are shown in Fig.~\ref{fig:figure_video}, see its caption for more details.

\begin{enumerate}

\vspace{0.5cm}

    \item Movie1
    shows the entire dynamics of phase separation, starting from a random uniformly distributed configuration at the desired packing fraction, $\phi=0.480$. The quench is done to Pe $=200$, well within the MIPS sector of the phase diagram. The video focuses distinctly  
    on the three dynamical regimes highlighted in Fig.~2 
    of the main text. The first regime of nucleation, from $t \sim 10^{-2}$ to $t \sim 5\times 10^{-2}$, is shown with  a rate of $\sim 5 
    \times 10^{-4}$ time units per second. As confirmed quantitatively in Fig.~\ref{fig:figure_video}(a), the number of clusters 
    grows in this first regime, and the structure factor, see Fig.~\ref{fig:structure-factor}, begins to develop a short-wavelength 
    peak, as a result of the formation of small clusters. Within the second regime, from $t \sim 5 \times 10^{-2}$ to 
    $t \sim 5 \times 10^{-1}$, the condensation of particles from the gaseous phase into the clusters, and the coalescence of 
    macroscopic clusters, contribute to the growth of the dense phase. This is corroborated in Fig.~\ref{fig:figure_video}(a), 
    where one observes the  decrease of both 
    the number of clusters and the total mass of the gaseous phase. This regime is shown in the movie with a rate of 
    $\sim 5 \times 10^{-2}$ time units per second. Finally, within the dynamical scaling regime, shown with $\sim 0.5$ time units per 
    second, the mass of the two phases is conserved, and the average cluster size grows due to merging of clusters (accompanied 
    by some clusters that evaporate and break, and later recombine) as shown in Fig.~\ref{fig:figure_video}(a).
    
\vspace{0.5cm}

    \item 
    Movie2
    displays the evolution of a system with $\phi=0.500$ and Pe $=100$, starting from an initial configuration engineered as a disk with uniform local hexatic order. Positions of particles inside the disks are extracted from a stationary configuration of a system with quasi-long-range hexatic order, evolved at the same Pe $=100$ but higher global surface fraction, which selects the right target binodal density for the dense phase in MIPS. The disk occupies half of the total area of the system, corresponding to the same amount we observe at stationarity for global $\phi=0.500$ (see main text). The free space around the disk is filled with randomly located particles. Constant global mass of the gaseous phase during evolution from such initial state, shown in Fig.~\ref{fig:figure_video}(b), confirms that the chosen partition of the system is the right one.
    The video demonstrates that the global 
    orientational order is progressively lost: the disk breaks into pieces of smaller size, as shown by the increasing number of hexatic clusters and bubbles in between  them in Fig.~\ref{fig:figure_video}(b).
   
   \vspace{0.5cm}
   
    \item 
    Movie3 presents the evolution of the same system starting now from an initial state with an ordered slab, built with the same procedure as the one used for the disk.
    
\end{enumerate}

\vspace{1cm}

 {
\section{II. Local density}}
\label{sec:snapshots}

\begin{figure}[t!]
\vspace{0.25cm}
\includegraphics[width=15cm]{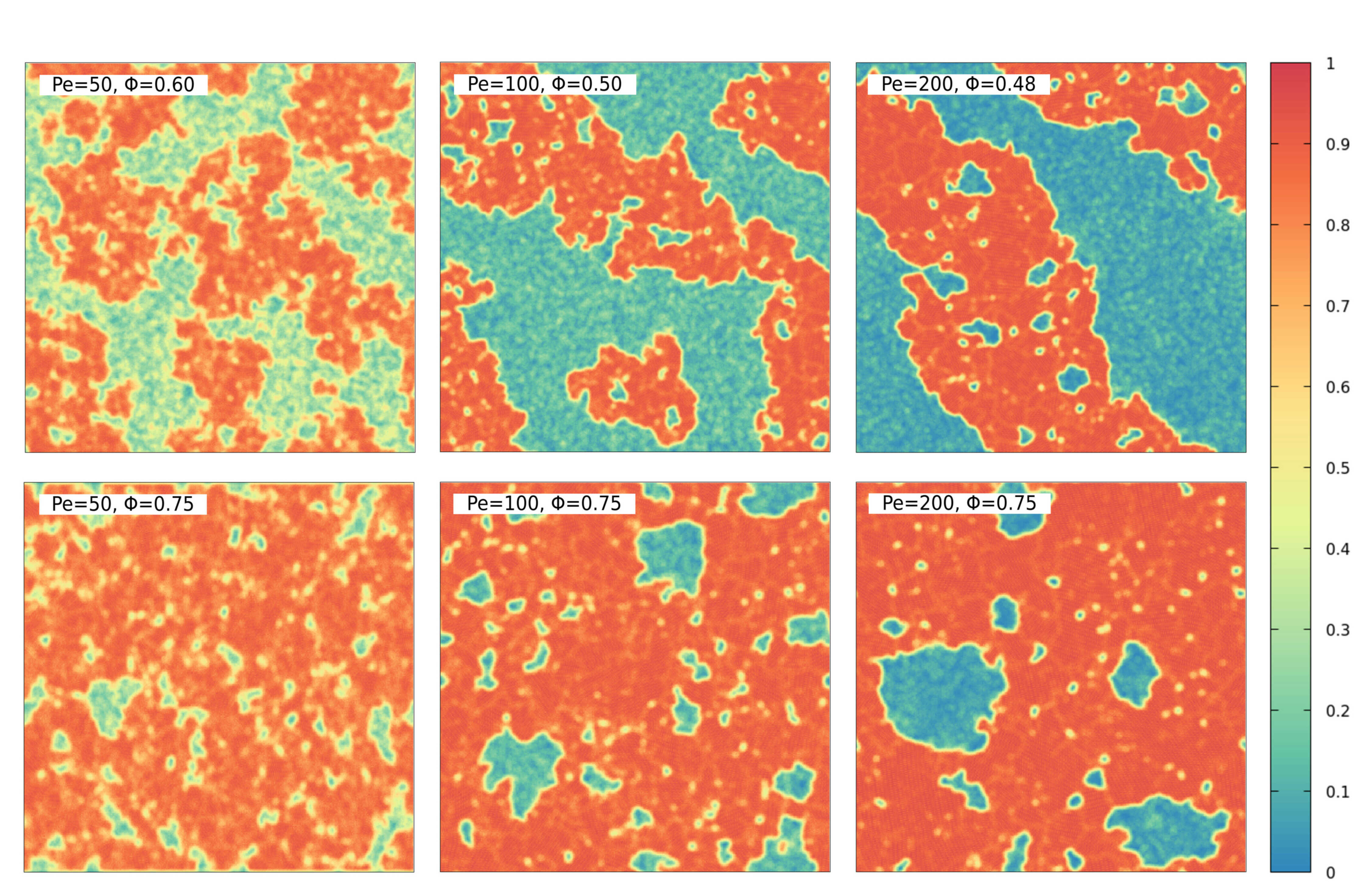}
\caption{Local density map of different steady-state configurations of systems with $N=512^2$ ABP  
using the color scale shown in the right bar. The parameters are given as labels within each panel.
}
\label{fig:snap-bubble}
\end{figure}

Figure~\ref{fig:snap-bubble} shows six snapshots in the form of heat maps of the local density, according to the
scale in the right vertical bar, running from close packed (red) to the dilute limit (blue). These snapshots correspond to steady-state configurations of $N=512^2$ ABPs for the different  values of $\phi$ and Pe indicated in the key. The panels are ordered in 
such a way that Pe increases from left to right. 
On the first row the system  phase separates into half dense and half dilute, 
while on the second row the packing fraction is just constant, $\phi=0.75$.
In all cases one clearly notices the phase separation between dense and dilute regions. Several other features can be 
noted as well. First, for increasing Pe, the density contrast between the two coexisting phases is more pronounced, 
and  thus the phase boundaries (in yellowish)  become sharper. 
Second, in all cases gas bubbles populate the dense (red) phase.
Third, the bubbles have different sizes and their characteristic average size increases with  Pe. Fourth, the density of the bubbles inside the dense phase seems  close to the one of the surrounding dilute phase.

\vspace{0.5cm}

\begin{figure}[t!]
    \vspace{0.25cm}
    \includegraphics[width=\textwidth]{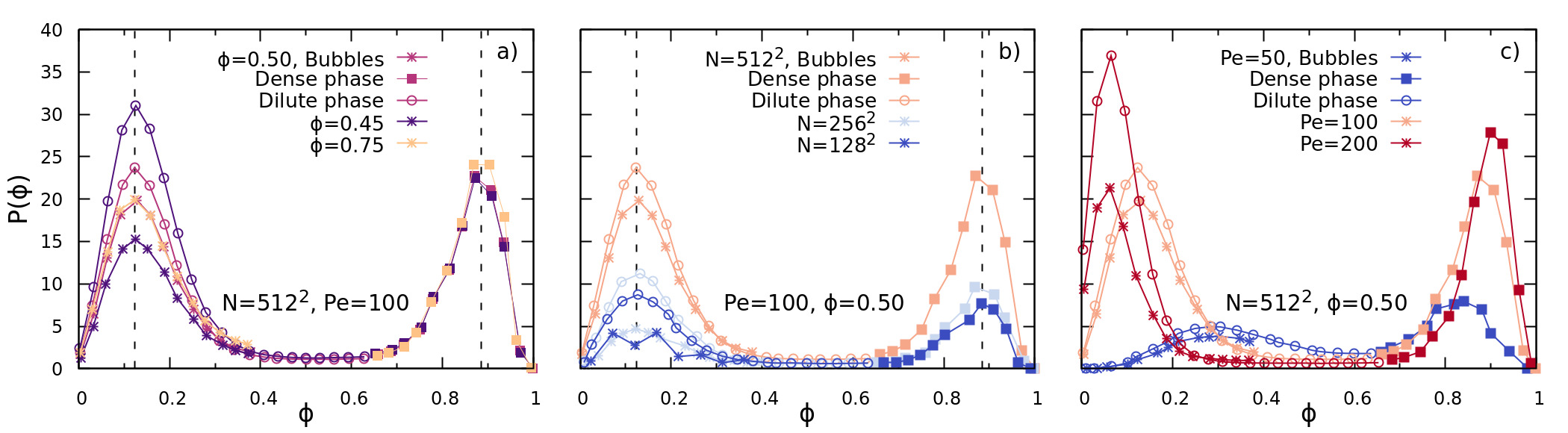}
    \caption{ {Probability distribution function of the local density of the bubbles (star symbols), dense (square symbols) and dilute (open circle symbols) phases for three different simulation parameters $N$, $\text{Pe}$ and $\phi$. The dotted vertical lines are a guide to the eye, and locate the peaks' positions. In each panel, the two parameters in the label are fixed, and the values of the other one are represented using different colors. As shown in the first two panels, varying the density (a) or the number of particles (b) does not affect the peaks' position of the distribution, but only their relative weight $P(\phi_{\text{peak}})$. In particular, the density values $\phi_{\text{peak}}$ for the dense phase and the bubbles are the same for each parameter choice. In panel (c) we show that the peaks move apart by increasing $\text{Pe}$, and the gap between the dense and dilute densities increases. For each $\text{Pe}$ value, the bubbles' and  dilute phase peaks coincide.} 
}
\label{fig:pdfbubble}
\end{figure}

 {
In order to quantify this last point,  in Fig. \ref{fig:pdfbubble}  we show the local density probability distribution function (pdf) of the dilute and dense phase, and of the bubbles, for different parameter values shown in the key. As expected, the density pdfs split in two modes in the MIPS regime, and the location of the low-density peak matches the one of the bubbles' density pdf, thus showing that the dilute phase and the bubbles share the same average density. 
}

\vspace{1cm}

\section{III. Dynamical scaling}
\label{sec:dynamical_scaling}

\begin{figure}[h!]
\includegraphics[width=10cm]{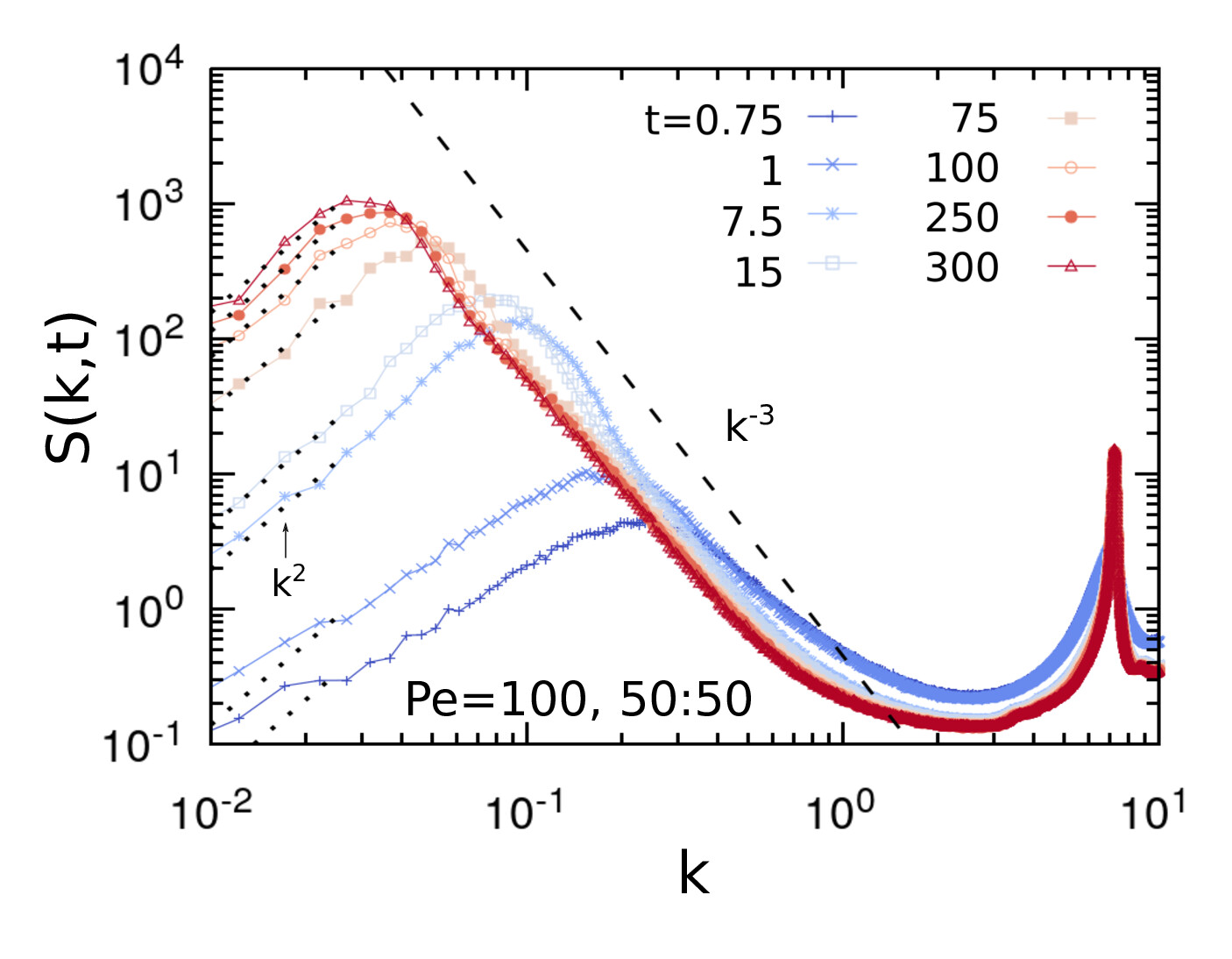}
\caption{Structure factor $S(k,t)$ calculated from simulations of $N=1024^2$ ABP at $\text{Pe}=100$ 
(using a 50:50 mixture) and for the different times after the quench reported in the key. The dotted lines are the $k^2$ growth 
for very small wave vector expected in the long time scaling limit.
The first peak is at $\hat k(t) = \pi/R(t)$ with $R(t)$ the typical growing length of the dense phase.
The dashed line $\sim k^{-3}$ is shown as a reference 
for the decay on the long-wavelength side of the peak, according to Porod's law. The second peak is 
located at  $k \sim 2\pi/\sigma_d$ and it does vary with time. }
\label{fig:structure-factor}
\end{figure}

The time-dependent  spherically averaged structure factor $S(k,t)$ (associated to the density-density correlations) 
at several times $t$ after the quench
is shown in Fig.~\ref{fig:structure-factor}. The increase in magnitude 
of the short wave-length 
peak as a function of time and its shift towards lower wave-vector values follow the large scale ordering kinetics of the system 
after being  quenched to high activity from a randomly disordered state (see the main text for details about the quench). 
The location of the first peak, $\hat k(t)$, corresponds to the characteristic length scale in the system, 
the dense cluster(s) mean size, by the relation $R(t)=\pi/\hat k(t)$.
The structure factor also shows a second peak at a time-independent wave-length, $k \sim 2\pi/\sigma_d$, 
that is related to the short-distance hexatic structure of the dense phase~\cite{linoSM}. 
 After the time scale associated to the nucleation of small aggregates from the homogeneous disordered phase, the position of this peak does not vary.  This time-independence suggests that the local structure of the system does not significantly change over the scale of the first neighbor shells, while the size of the dense clusters grows.
The large-scale behavior, in between the first and second peak,  is in agreement with Porod's law, $S(k,t) \sim k^{-(d+1)}$, with $d=2$ the dimensionality of the system, as expected for segregated systems with smooth interfaces between the phases~\cite{braySM}.

\begin{figure}[h!]
\vspace{0.25cm}
\includegraphics[width=18cm]{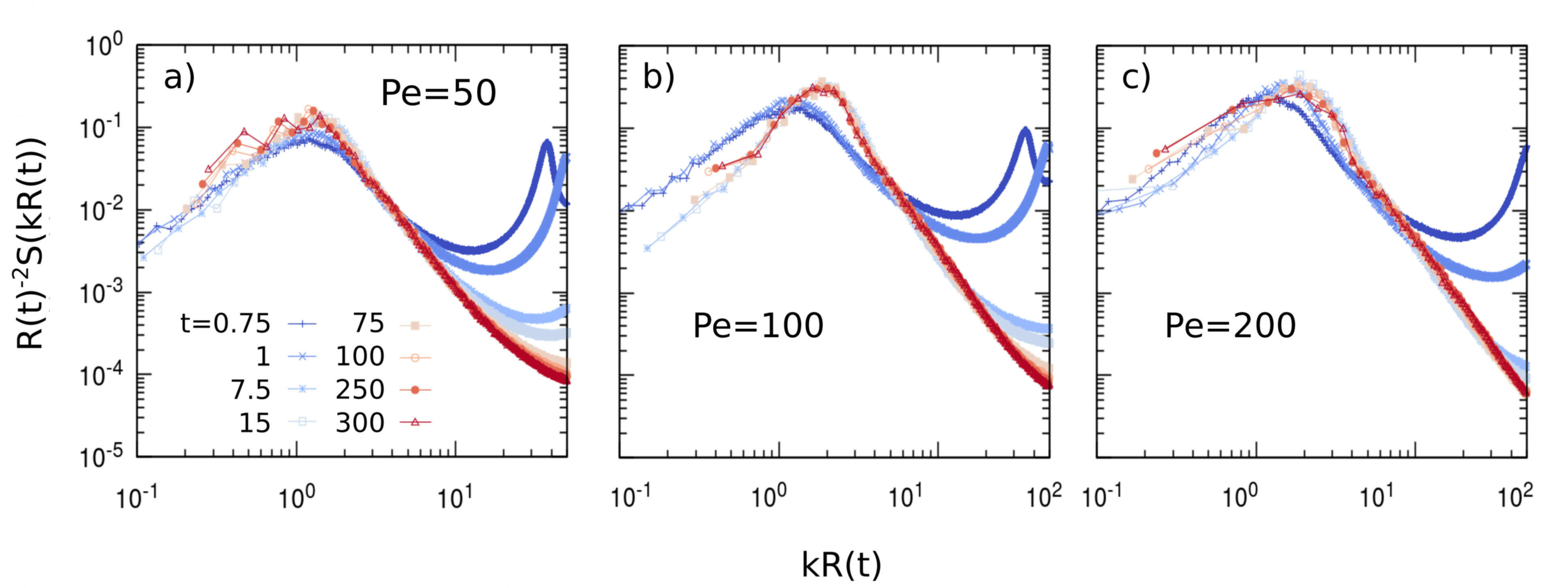}
\caption{Scaling of the structure factor according to the form in Eq.~(\ref{eq:dyn-scaling}). Data obtained  
from simulations of a system of $N=1024^2$ ABP with equal fraction of the system in the dilute and dense phase. 
The measuring times are reported in the key in (a), and the different $\text{Pe}$ values are indicated as labels. 
The data points fall on top of each other for $t>7.5$ in all cases, showing that dynamical scaling holds at all Pe.}
\label{fig:dynamic-scaling}
\end{figure}

Numerical simulations~\cite{stenSM} suggested that MIPS in ABP verifies dynamical scaling~\cite{braySM}, meaning that at the late stages of phase separation, the evolution of the structure factor is characterized by a single length scale:
\begin{equation}
S(k,t) = R(t)^{d} \, \mathcal{F}(kR(t)) 
\; .
\label{eq:dyn-scaling}
\end{equation}
We extensively tested this hypothesis in our simulations with $N=1024^2$ particles. We used systems made of an equal fraction of dense and dilute phases (50:50 fraction), for  different $\text{Pe}$ values. In Fig.~\ref{fig:dynamic-scaling} we show scaled data   
using Eq.~(\ref{eq:dyn-scaling}) and $R(t)$ extracted from the numerical
data (see Sec. IV A below).
The data collapse at late times ($t\gtrsim 7.5$ in units of $1/D_{\theta}$) allows us to define the ``scaling regime'', during which the dynamics fulfills Eq.~(\ref{eq:dyn-scaling}), as described in the main text. At early times dynamical scaling is not satisfied, meaning that several length scales are present in the initial growing kinetics. Indeed, we identified two other regimes 
before scaling sets in. In these two earlier  regimes, the growth is lead by two different mechanisms
described in the main text.

Finally, Furukawa showed that for conserved scalar order parameter  dynamics (model B continuous field theory)
the small $k$ behavior of the scaling function $\mathcal{F}(kR(t)=x) \sim x^2$ grows from zero as $k^2$, if thermal fluctuations are effective~\cite{furuSM}. The small wave-vector structure factor of the ABP behaves in this way at long times, as 
shown with the dotted lines added close to the data at different times in Fig.~\ref{fig:structure-factor}. Already at 
$t=15$ the long-time behaviour is attained. (Note that in the particle system 
$S({\mathbf 0},t)=N$.
The first data-points in Figs.~\ref{fig:structure-factor} 
correspond to the first wave-vector available, with components $2\pi/L$.)

\section{IV. Numerical methods}
\label{sec:methods}

In this Section we explain the numerical techniques that we used to identify the dense phase, or droplet, the hexatic domains  and the gas bubbles.

\subsection{A. Dense phase size}
\label{subsec:droplet}

In Sec. III we used the first peak of the structure factor to identify the growing length of the 
dense phase. An equivalent measure of the single characteristic length in the scaling regime
could be given by the 
first moment $\bar{k}$ of the structure factor, over the range $[0 \! : \! \pi]$, 
\begin{equation}
R(t)=\pi/\bar{k}(t) {\bm .}
\end{equation}
This quantity also allowed us to analyze the growth rate of particle clusters, the dense phase, and compare 
it to previous results from continuum theories.

Similar results as the ones resulting from the structure factor analysis arise from a more explicit measurement of the average size of particle aggregates, obtained by applying a standard DBSCAN algorithm to the positions of the disks~\cite{esterSM}. DBSCAN is a clustering algorithm, which distributes points into clusters according to the local point density. We shortly outline hereafter the fundamental rules of the algorithm, in order to justify our choice of parameters.
\begin{itemize}
 \item Given that two points are neighbors if their distance is less than a given extent $\varepsilon$, a point is a ``core point'' if it 
 has at least $n_{\rm min}$ neighbors;
 \item any two core points connected through a path in the neighbors network belong to the same cluster, together with their neighbors;
 \item points which are not cores and are not reachable from a core do not belong to any cluster.
\end{itemize}
We used $\varepsilon=1.5 \, \sigma_d$ and $n_{\rm min}=6$ for a successful identification of the clusters (relying on previous results at  different Pe values~\cite{linoSM}). Since particles in the dense phase are locally arranged on a hexagonal lattice with local surface fraction ranging from $\phi_{\rm loc}\sim 0.800$ at Pe $=50$ to $\phi_{\rm loc}\sim0.900$ at Pe $=200$, a circle of radius $\varepsilon=1.5 \, \sigma_d$ encloses the first shell of neighbors. 

\begin{figure}[h!]
\vspace{0.25cm}
\includegraphics[height=0.4\textwidth]{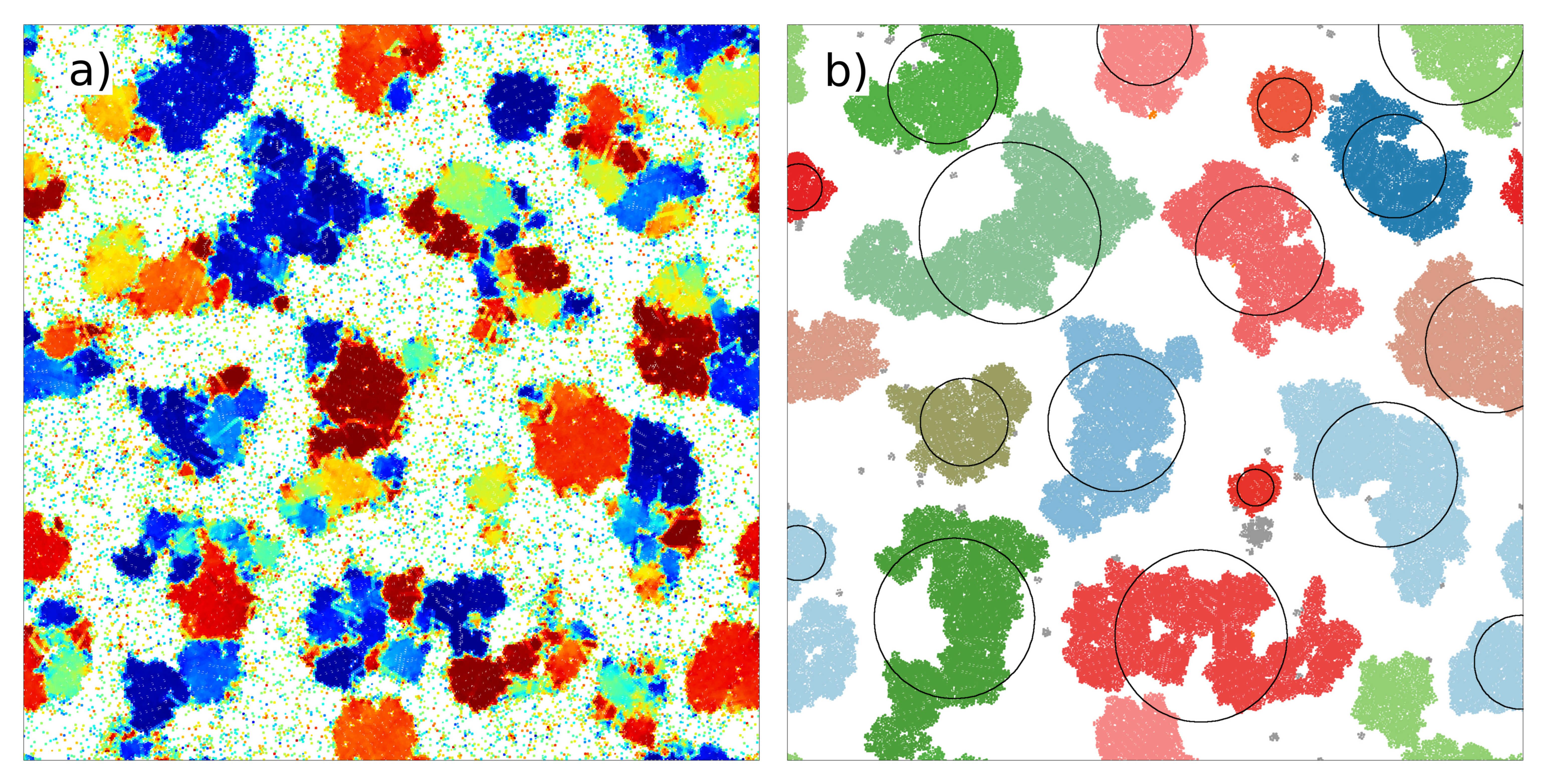}
\caption{(a) A configuration drawn using colors that represent the local orientational order. 
(b) Identification of the dense drops in (a) using DBSCAN (see text for the details). The same color is used for particles belonging to the same cluster, while isolated particles are not printed. Grey is used for small clusters, which have been discarded from the average. Black circles are centered at the center of mass of each cluster and have a radius equal to the radius of gyration of the same cluster.}
\label{fig:dbscan}
\end{figure}

An example of the performance of this algorithm is shown in Fig.~\ref{fig:dbscan}. In (a) we show a configuration using the representation in which we paint each particle with a color associated to its hexatic order $\psi_{6,i}$ (see Fig. 1 in the main text). As in~\cite{linoSM}, red indicates maximal projection on the averaged orientation of the full sample in the $\psi_6$ space, and blue the maximal projection in the opposite direction, with a usual color scale in between these two extremes. One clearly observes  in Fig.~\ref{fig:dbscan}(a) phase separation between very dilute regions and rather dense ones composed of patches with different orientational order. In panel (b) we show the outcome of the use of DBSCAN to identify the clusters (each cluster is shown with a different color). 
The size of the clusters is then calculated from their radius of gyration, 
\begin{equation}
r^i_G= \sqrt{s_i^{-1} \sum_{j=1}^{s_i} ({\bm r}_j-{\bm r}^i_{\rm cm})^2} 
\end{equation}
where $s_i$ is the number of particles in  cluster $i$, ${\bm r}_j$ the positions of particles belonging to the cluster, and ${\bm r}_{\rm cm}^i$ its center of mass. A black circle, centered at the center of mass and with radius $r^i_G$, is drawn on each cluster.
We then monitor the average radius of gyration $R_G(t)$ obtained after averaging over all the clusters: $R_G(t)=N_C^{-1}\sum_i r^i_G$, $N_C$ being the total number of clusters in the system.

Although the algorithm is quite reliable on the clusters' identification, it is however not able to distinguish between macroscopic 
clusters leading the coarsening and very small aggregates, the latter being not stable since they continuously arise and evaporate 
within a timescale of a few simulation time-steps. In order to avoid their impact on the system averages, 
we discard, as a ``rule of thumb'', all the clusters containing less than $5\%$ particles with respect to the largest one.

\subsection{B. Hexatic domains}

Large dense clusters do not always have hexatic order to the scale of their whole size~\cite{linoSM} but, instead, 
they are arranged in a ``polycrystal'' or ``mosaic'' of hexatically ordered domains, with almost uniform local hexatic parameter $\psi_{6,i} = N_i^{-1} \sum_j^{N_i} e^{i6\theta_{ij}}$ in their interior. We describe below two methods used to measure the average size of these domains, which yield consistent results. We either use a {\it clustering by argument} approach or a {\it clustering by gradient} one, to first identify hexatic domains, and then measure their size. Both methods are  applied to the particles in the dense phase only, which are previously selected by the application of DBSCAN, as described in Sec. IV A.

\begin{enumerate}

\item {\it Clustering by argument.}

Within this approach we discretize the range $[0 \! : \! 2\pi]$ of the argument of $\psi_{6,i}$ into $n$ bins and we split the system accordingly. We then apply the DBSCAN algorithm to each part of the system separately and we discard, according to the 
rule of thumb introduced in Sec. IV A, the clusters with less than $5\%$ particles of the largest one. The two steps are shown in Fig.~\ref{fig:hexatic-arg}. We verified that the results depend only weakly on $n$ within the range $n=4, \dots, 10$, 
as shown in Fig.~\ref{fig:comparison}(a). We used $n=6$ for all the measures presented in this work. 

 \begin{figure}[h!]
\vspace{0.25cm}
\includegraphics[height=0.41\textwidth]{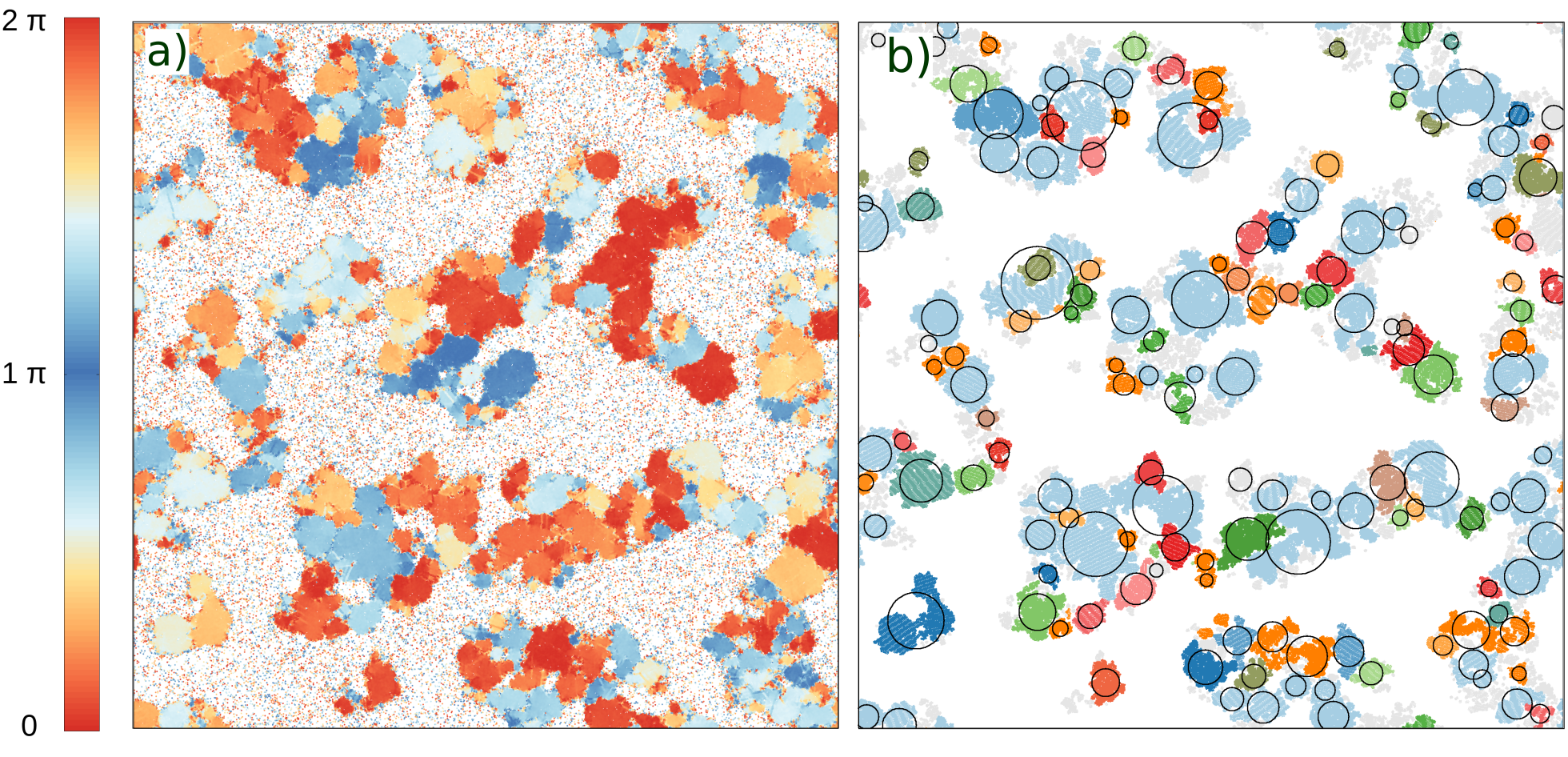}
\caption{Identification of hexatic domains using the argument of the local hexatic order parameter $\psi_{6_i}$. (a)  {Color map of ${\rm Arg}(\psi_{6,j})$}. (b) Same system, split according to the binning of the hexatic orientation. Black circles are centered at the center of mass of each hexatic domain and have a radius equal to their radius of gyration.}
\label{fig:hexatic-arg}
\end{figure} 


\item{\it Clustering by gradient.}

As an alternative approach to separate hexatic domains, we developed a criterion based on the spatial gradient of the argument of the local hexatic parameter. We first coarse-grain the local hexatic order parameter on a square grid of spacing $d=5\sigma_d$, being such coarse-graining length smaller than the typical size on any hexatic domain. Then we calculate the gradient of the coarse-grained ${\rm Arg}(\psi_6^{i,j})$. We then associate to each grid point a `0' or `1' if $\lvert \nabla{\rm Arg}(\psi_6^{i,j}) \rvert$ is larger or smaller than a certain threshold, which here we fixed  to ${\rm th}_{grad}=0.2$ (in units of $1/\sigma_d$), which is approximately the 10\% of the typical range of the gradient modulus. Grid points labeled with a `1' are considered to belong to an hexatic domain.
We then apply a DBSCAN algorithm to the grid points with  $\lvert \nabla{\rm Arg}(\psi_6^{i,j}) \rvert<{\rm th}_{grad}$ in order to identify the hexatic domains. Compared to the previous DBSCAN algorithm, in this case we select only the first eight neighbors for each grid point, and $n_{\rm min}=4$. The radius of gyration is evaluated from the positions of the grid points.
Such approach is graphically summarized in Fig.~\ref{fig:hexatic-grad}.

\end{enumerate}

 \begin{figure}[h!]
\vspace{0.25cm}
\includegraphics[height=0.4\textwidth]{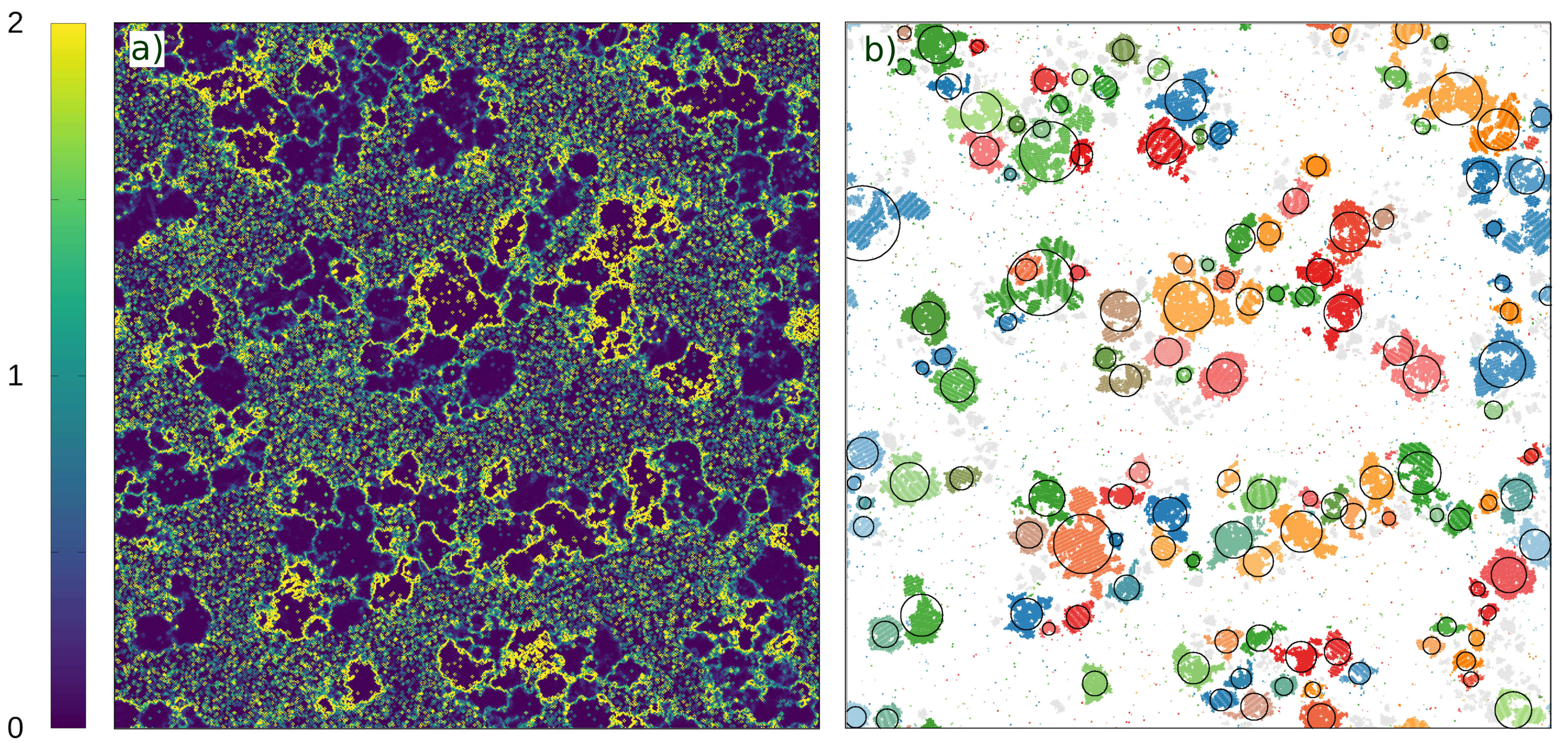}
\caption{Identification of hexatic domains using the spatial gradient of ${\rm Arg}(\psi_6)$. (a) Gradient field computed over a square grid of side $\sigma_d$. (b) Hexatic domains, as identified from the lattice DBSCAN applied to regions in (a) where $\lvert \nabla({\rm Arg(\psi_6)})\rvert<0.2$.}
\label{fig:hexatic-grad}
\end{figure}

In Fig.~\ref{fig:comparison}(b) we compare the two methods described above to identify hexatic domains. We plot the averaged radius of gyration of 
the hexatic domains found with each technique and we show that the results are consistent. In the main text we present results found with the clustering by
argument method only, with $n=6$.

\begin{figure}[h!]
\vspace{0.23cm}
\includegraphics[height=0.35\textwidth]{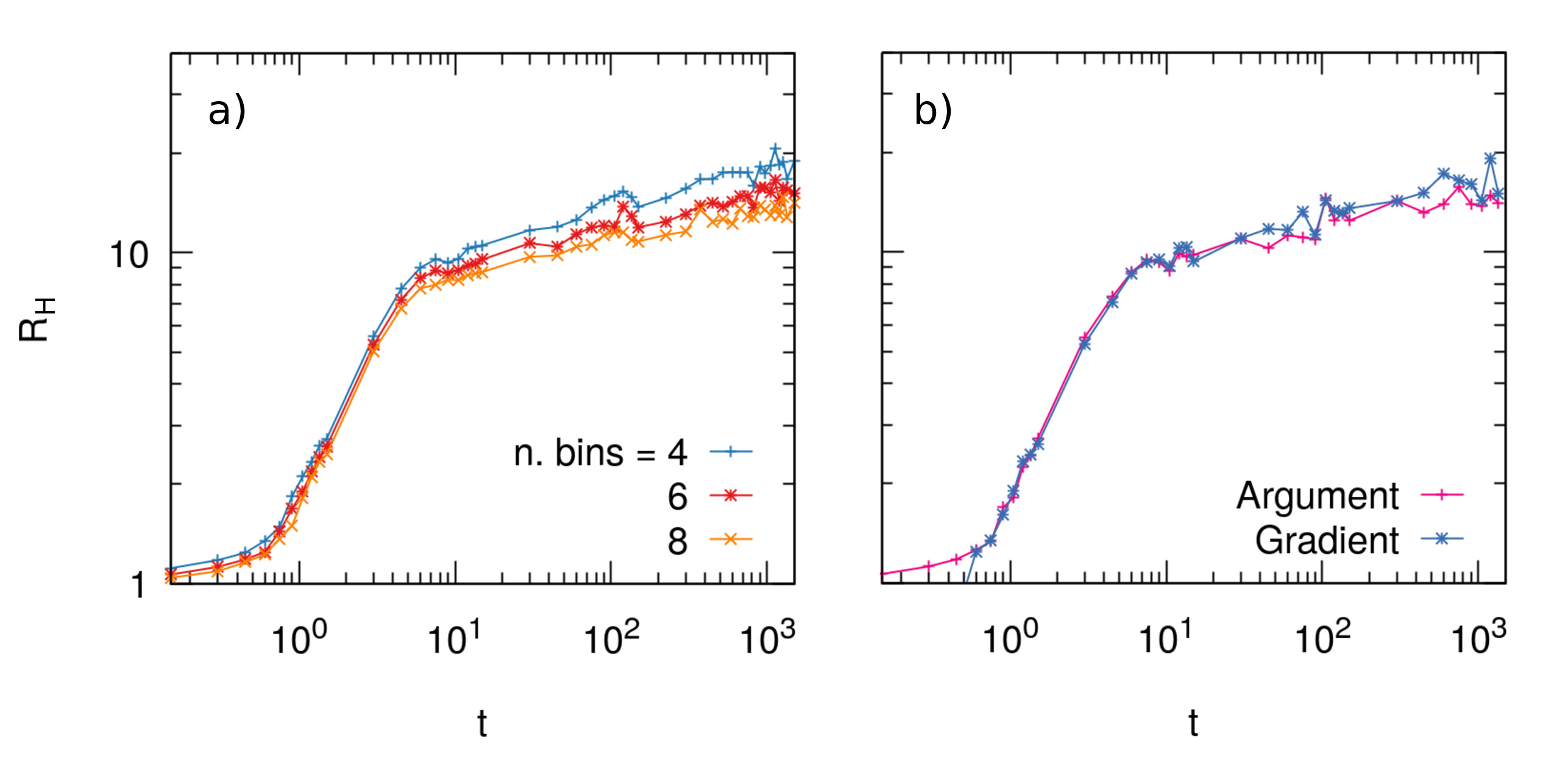}
\caption{The average size of clusters with the same hexatic order  as a function of time.
(a) Clustering by argument method:  comparison between results obtained from a single simulation, using different number of bins  for the discretization of the argument. (b) Comparison of results obtained using the clustering by argument (with $n=6$) and clustering by gradient methods.}
\label{fig:comparison}
\end{figure}



\subsection{C. Bubble identification}

\begin{figure}[t!]
\vspace{0.25cm}
\includegraphics[height=0.35\textwidth]{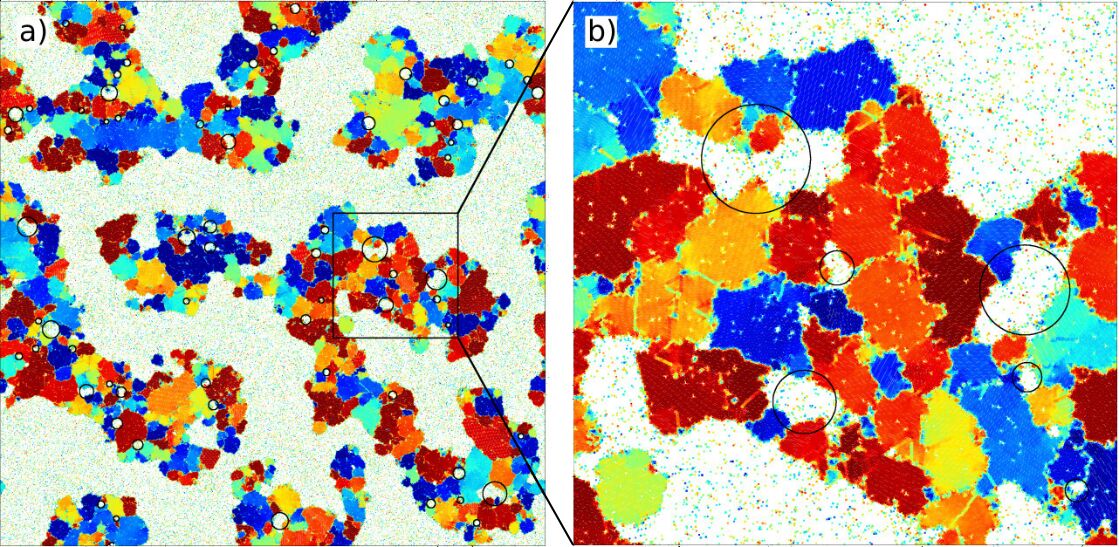}
\caption{(a) Hexatic map of a system with $N=1024^2$ particles, Pe $=200$ and $\phi=0.480$. Black circles are centered at the center of mass of each bubble, and have radius equal to their radius of gyration. (b) Zoom over the black squared area in panel~(a).
}
\label{fig:bubble-algo}
\end{figure}

In order to identify the bubbles we first coarse-grain the system's local surface fraction over the length $d=5\, \sigma_d$. 
Specifically, we divide the system area into a grid of square cells of linear size $d$, we set to $1$ all the cells with average surface fraction $\phi_i$ below a threshold ${\rm th}_{\phi}=0.65$, and to $0$ the ones for which $\phi_i>{\rm th}_{\phi}$. The value of ${\rm th}_{\phi}$ has been chosen to trace the dense droplets contour into the grid as accurately as possible. Bubbles are then identified by means of the lattice-DBSCAN described above, as illustrated in Fig.~\ref{fig:bubble-algo}. The coarse-graining length $d$ we used is large enough to cut off all the small bubbles of size of the order of $\sigma_d$. The latter arise from extended defects inside the dense phase (particularly across the boundaries between different hexatic domains), and are not relevant to the large-scale phenomenon we aim to address. Finally, we also discard all the bubbles which span the whole system along either the $x$ or the $y$ axis, constituting the gaseous phase of the segregated system.

 {
\section{V. Hexatic order: growing length and dynamic scaling}
\label{sec:hexatic-growth}
}

 {In this Section we further investigate the growth of hexatic order.}

\begin{figure}[t!]
    \vspace{0.25cm}
    \includegraphics[width=0.7\textwidth]{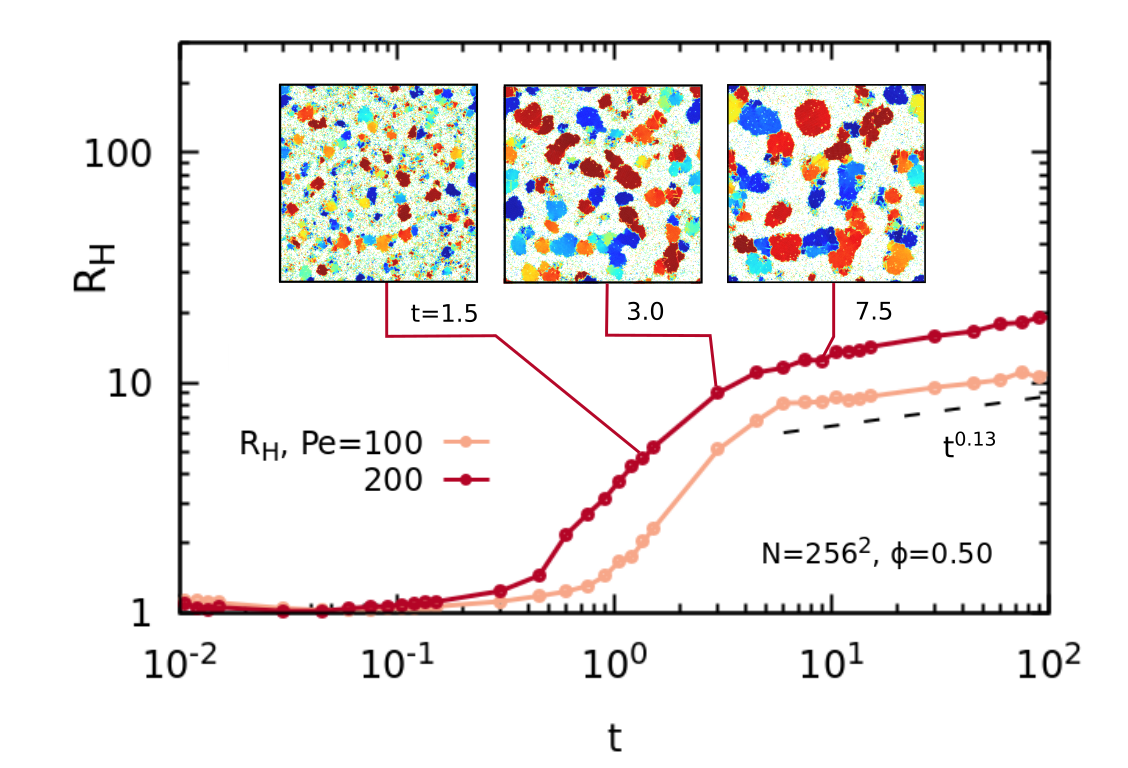}
    \caption{ {The time dependence of the averaged hexatic length together with three representative snapshots
    at times close to the crossover from aggregation/coagulation to the scaling regime.} 
    }
\label{fig:hex_length}
\end{figure}

 {
First, we plot in Fig.~\ref{fig:hex_length} the averaged hexatic order length scale as a function of time Pe = 100, 200 and 
$\phi =0.5$. We 
accompany the graph with three snapshots of configurations at selected times around the 
crossover from the $t$ (aggregation/coagulation) to the $t^{0.13}$ (scaling regime) algebraic growths.
}

 {Next,}
we verify, by means of a spectral analysis of the modulus of the local hexatic order parameter, that the eventually arrested 
hexatic order growth satisfies the dynamical scaling hypothesis over the same time regime for which the local density does, as shown in 
Sec. III. This is the regime that we called scaling in the main text and 
in which the hexatic patches grow in time very slowly, with a power law with exponent that we estimated to be $\sim 0.13$ 
for Pe = 100 and half and half dense and dilute components but could be slightly different for these or other parameters.

Concretely, we computed the following quantity
\begin{equation}
    \label{eq:hex_str_fact}
    S_H(\mathbf{k}) = \frac{1}{N} \langle |\psi_6|_{\mathbf k} \ |\psi_6|_{-\mathbf k} \rangle {\rm ,}
\end{equation}
where $|\psi_6|_{\mathbf k}$ is the Fourier transform of the modulus of the local hexatic order parameter, 
defined for each particle as described in the main text.
\begin{figure}[t!]
    \vspace{0.25cm}
    \includegraphics[width=\textwidth]{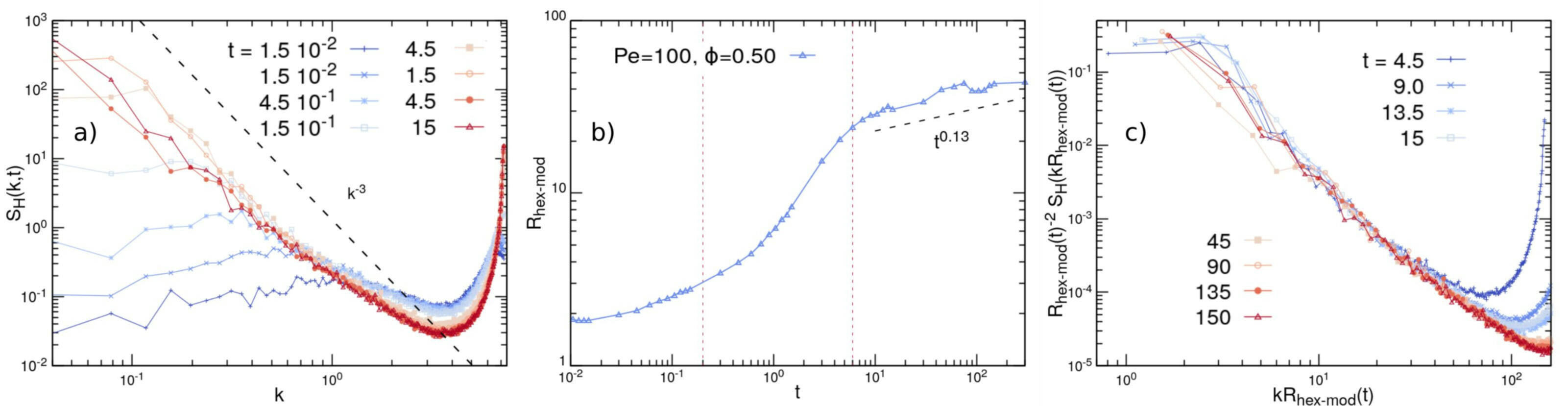}
    \caption{(a) Hexatic modulus structure factor, as defined in Eq.~(\ref{eq:hex_str_fact}), in a system with  Pe $=100$ and $\phi=0.500$. (b) Average hexatic size, computed as $\pi/\bar{k}$ from the first moment $\bar{k}$ of $S_H(k,t)$. The dashed black line $\sim t^{0.13}$ is shown 
    next to the data. It is the growth law found from the measurement of the averaged radius of gyration of the hexatic patches presented in the main text. (c) Scaled hexatic structure factor following a scale-free shape.
    }
\label{fig:hex_scaling}
\end{figure}
The results, for the spherically averaged quantity, are shown in Fig.~\ref{fig:hex_scaling}(a) for Pe $=100$, $\phi=0.500$ and different times spanning the entire dynamical range, from the disordered initial configuration to stationarity. Alongside the high wave-vector peak at $k \sim 2\pi$, which is related to the structure of the first neighbor shells, the hexatic modulus structure factor develops a low wave-vector peak. This represent hexatic domains and its location is shifted, from $k \sim 0.1$ at the early growing stages, to the left as time increases and the hexatic patches grow, reaching $k \sim 0.01$ at the beginning of the last stationary regime. The inverse of the first moment of $S_H$, which we call $R_{\rm hex-mod}(t)$, is shown Fig.~\ref{fig:hex_scaling}(b). It consistently shows the hexatic coarsening, and it very well agrees with our results on the power-law growth in the scaling regime, obtained from the analysis of the averaged gyration radius of the micro-domains.
The decay of $S_H$ from the hexatic peak towards lower wave-vectors satisfies the Porod's law (see Sec. III) with $d=2$, which allows us to formulate the right scaling hypothesis. Figure~\ref{fig:hex_scaling}(c) demonstrates the scaling of the structure factor, using the typical length-scale $R_{\rm hex-mod}(t)$.

 {
\section{VI. Equilibrium Hexatic-Liquid coexistence}
\label{sec:equilibrium}
}

\begin{figure}[h!]
    \vspace{0.25cm}
    \includegraphics[width=\textwidth]{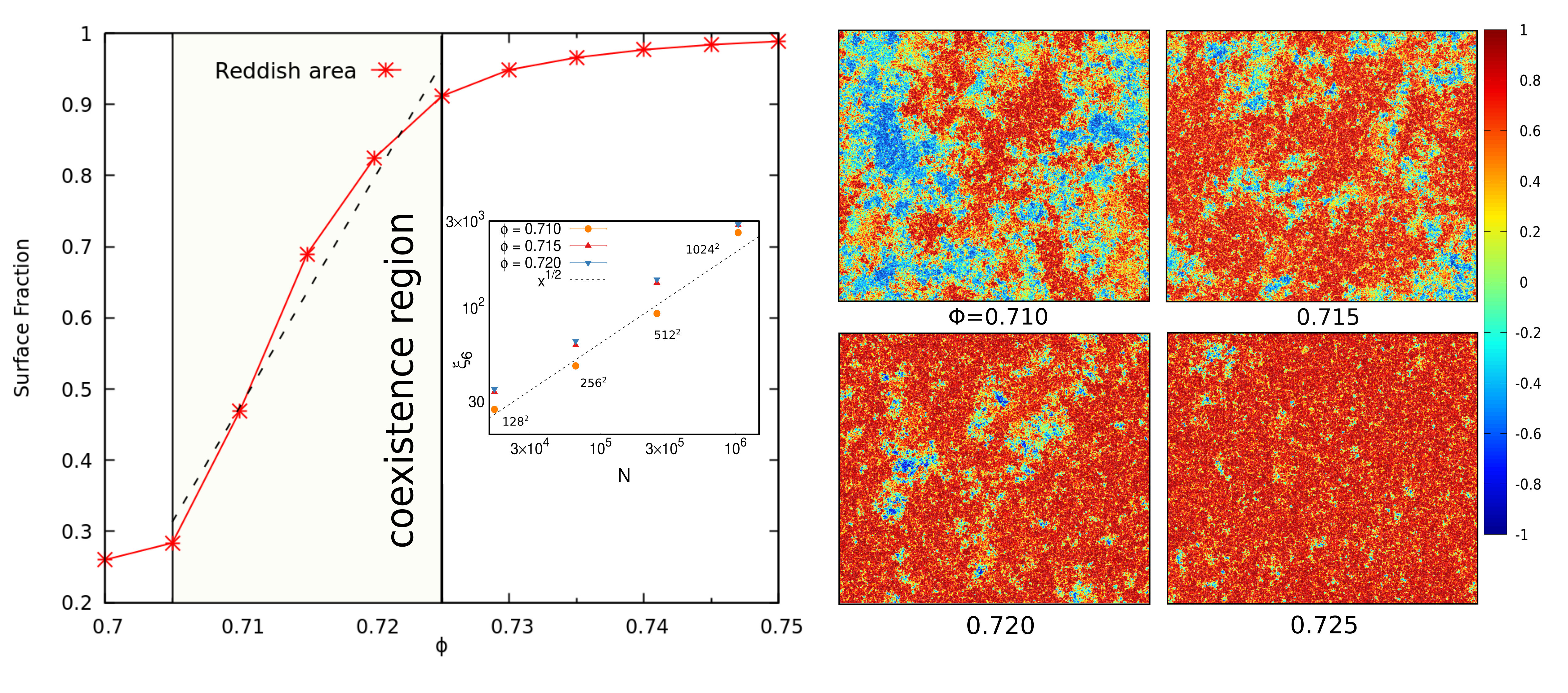}
    \caption{ {Fraction of the total surface of the system occupied by the hexatic phase as a function of the packing fraction across the equilibrium liquid-hexatic transition (left). In the coexistence region, indicated by two vertical lines, the evolution of the area fraction occupied by the hexatic as the density of the system increases, is compatible with a linear growth (shown by the dotted line). In the inset we show the hexatic correlation length $\xi_6$ as extracted from the spatial decay of  orientational correlations in systems of different size $N$ and at different packing fractions indicated in the key.  On the right we illustrate the growth of the hexatic area in the coexistence region with four snapshots showing how the reddish area corresponding to the hexatic phase grows at the packing fraction increases.} 
    }
\label{fig:EqCoexistence}
\end{figure}

 {
In this Section we provide further details on the nature of the hexatic-liquid coexistence in equilibrium, 
which differ from the non-equilibrium coexistence triggered by MIPS at high activity. 
After constructing a Voronoi tessellation out of configurations of our model at Pe = 0, we computed the ratio between the area covered by the sum of all the Voronoi cells for which the projection of $\psi_6$ along the mean orientation of the sample is $>0.4$. In such a way, we  tracked the growth of the fraction of the system occupied by the hexatic phase. The results of this analysis are shown in Fig. \ref{fig:EqCoexistence}. The snapshots on the right show the map of the projection of the local hexatic order parameter $\psi_6$  in the direction of its global average for four densities across the coexistence region at Pe = 0. 
As shown in the left panel and illustrated by the hexatic maps, the hexatic  is dominated by the mean orientation  (reddish area in the snapshots), and its extent grows across the  coexistence region as the density increases. The growth of the hexatic in the coexistence regime is compatible with a linear growth: as the density increases, the fraction of the system belonging to the dense hexatic phase increases proportionally, and thus the area covered by the hexatic. Note that this behavior strongly differs from the MIPS scenario, where the dense phase is made of a mosaic of patches with different orientation (different color in the hexatic maps, not a single reddish one) whose  size remains constant all along the MIPS coexistence region. 
}

 {
In the inset we display the size dependence of the typical length of the hexatically ordered zone $\xi_6$ extracted from the decay of the hexatic correlation function $g_6(r) = \langle \psi_6(0) \psi_6^*(r) \rangle / \langle |\psi_6(0)|^2  \rangle$. Its dependence on the system size is close to $\sqrt{N}$ for the three packing fractions shown. Such behavior is expected as the length scale associated to hexatic order grows with the size of the system and the fraction of it belonging to each phase. However, this contrasts with the behavior across MIPS where the hexatic length scale remains finite and constant independently of the system size.  
}

\begin{thebibliography}{99}
\bibitem{MarchettiRev}
Marchetti M.C., Joanny J.F., Ramaswamy S., Liverpool T.B., Prost J., Rao M., Simha R.A., Rev. Mod. Phys., \textbf{85}, 1143 (2013).
\bibitem{WinklerRev}
Shaebani M.R., Wysocki A.,Winkler R.G., Gompper G., Rieger H., Nat. Rev. Phys. \textbf{2}, 181 (2020).
\bibitem{CatesRev}
Cates M.E., Tailleur J., Annu. Rev. Cond. Matt. Phys., \textbf{6}, (2015).
\bibitem{Romanczuk2012}
Romanczuk P., B{\"a}r M., Ebeling W., Lindner B., Schimansky-Geier L., Eur. Phys. J. Spec. Topics, \textbf{202}, (2012)).
\bibitem{Bialke2012cryst}
Bialk{\'e} J., Speck T., L{\"o}wen H., Phys. Rev. Lett., \textbf{108}, 168301, (2012).
\bibitem{Fily2012}
Fily Y., Marchetti M.C., Phys. Rev. Lett., \textbf{108}, 235702, (2012).
\bibitem{Stenhammar2014}
Stenhammar J., Marenduzzo D., Allen R.J., Cates M.E., Soft Matter, \textbf{10}, 1489, (2014).
\bibitem{Joan}
Levis D., Codina J., Pagonabarraga I., Soft Matter, \textbf{13}, 8113, (2017).
\bibitem{Speck15}
Speck T., Menzel A.M., Bialk\'e J., L\"owen H., J. Chem. Phys., \textbf{142}, 224109 (2015).
\bibitem{Redner13}
Redner G.S., Hagan M.F., Baskaran A., Phys. Rev. Lett., \textbf{110}, 055701, (2013).
\bibitem{PRLino}
Digregorio P., Levis D., Suma A., Cugliandolo L.F., Gonnella G., Pagonabarraga I., Phys. Rev. Lett., \textbf{121}, 098003, (2018).
\bibitem{KKK}
Klamser J.U., Kapfer S.C., Krauth W., Nat. Comm., \textbf{9}, 5045, (2018).
\bibitem{defectsLino}
Digregorio P., Levis D., Cugliandolo L.F., Gonnella G., Pagonabarraga I., arXiv:1911.06366 (2019).
\bibitem{PaliwalDijkstra}
Paliwal S., Dijkstra M., Phys. Rev. Res., \textbf{2}, 012013, (2020).
\bibitem{BennoLowen}
Mandal S., Liebchen B., L{\"o}wen H., Phys. Rev. Lett., \textbf{123}, 228001, (2019).
\bibitem{CapriniVelocities}
Caprini L., Marconi U.M.B., Puglisi A., Phys. Rev. Lett., \textbf{124}, 078001, (2020).
\bibitem{Stenhammar13}
Stenhammar J., Tiribocchi A., Allen R.J., Marenduzzo D., Cates M.E., Phys. Rev. Lett., \textbf{111}, 145702, (2013).
\bibitem{wittkowski2014}
Wittkowski R., Tiribocchi A., Stenhammar J., Allen R.J., Marenduzzo D., Cates M.E., Nature Comm., \textbf{5}, 1, (2014).
\bibitem{Speck14}
Speck T., Bialk\'e J., Menzel A.M., L\"owen H., Phys. Rev. Lett., \textbf{112}, 218304, (2014).
\bibitem{Tjhung18}
Tjhung E.,Nardini C., Cates M.E., Phys. Rev. X, \textbf{8}, 031080, (2018).
\bibitem{SM}
See Supplemental Material for further details.
\bibitem{BennoCluster}
Liebchen B., Marenduzzo D., Pagonabarraga I., Cates M.E., Phys. Rev. Lett., \textbf{115}, 258301, (2015).
\bibitem{BennoCluster2}
Liebchen B., Marenduzzo D., Cates M.E., Phys. Rev. Lett., \textbf{118}, 268001, (2017).
\bibitem{ChantalFrenkel}
Mognetti B.M., {\v{S}}ari{\'c} A.,Angioletti-Uberti S., Cacciuto A., Valeriani C., Frenkel D., Phys. Rev. Lett., \textbf{111}, 245702, (2013).
\bibitem{PacoChantal}
Alarc{\'o}n F., Valeriani C., Pagonabarraga I., Ignacio, Soft Matter, \textbf{13}, 814, (2017).
\bibitem{SolonChate}
Solon A.P., Chat{\'e} H., Tailleur J., Phys. Rev. Lett., \textbf{114}, 068101, (2015).
\bibitem{BennoPRL}
Liebchen B., Levis D., Phys. Rev. Lett., \textbf{119}, 058002, (2017).
\bibitem{GiomiPierce}
You Z., Pearce D.J.G., Sengupta A., Giomi L., Phys. Rev. X, \textbf{8}, 031065, (2018).
\bibitem{note1}
Here, the MIPS critical point is located at (Pe, $\phi)\approx(32,0.6)$.
\bibitem{proceeding}
Digregorio P., Levis D., Suma A., Cugliandolo L.F., Gonnella G., Pagonabarraga I., J. Phys.: Conf. Series, \textbf{1163}, 012073, (2019).
\bibitem{BrayRev}
Bray A.J., Adv. in Phys., \textbf{51}, 481, (2002).
\bibitem{Leyvraz03}
Leyvraz F., Phys. Rep., \textbf{383}, 95, (2003).
\bibitem{Cross93}
Cross M.C., Hohenberg P.C., Rev. Mod. Phys., \textbf{65}, 851, (1993).
\bibitem{Vega05}
Vega D.A., Harrison C.K., Angelescu D.E., Trawick M.L., Huse D.A., Chaikin P.M., Register R.A., Phys. Rev. E, \textbf{71}, 061803, (2005).
\bibitem{Kim19}
Hajibabaei A., Kim K.S., Phys. Rev. E \textbf{99}, 022145 (2019).
\bibitem{Pica20}
Li Y.-W., Pica Ciamarra M., Phys. Rev. Lett. \textbf{124}, 218002
\bibitem{Cecile2012}
Theurkauff I., Cottin-Bizonne C., Palacci J., Ybert C., Bocquet L., Phys. Rev. Lett., \textbf{108}, 268303, (2012).
\bibitem{Schaller2010}
Schaller V., Weber C., Semmrich C., Frey E., Bausch A.R., Nature, \textbf{467}, 73, (2010).
\bibitem{JulicherRev2019}
Weber C.A., Zwicker D., J{\"u}licher F., Lee C.F., Rep. Prog. Phys., \textbf{82}, 064601, (2019).
\bibitem{Tartaglia18}
Tartaglia A., Cugliandolo L.F., Picco M., J. Stat. Mech., \textbf{2018}, (2018).
\end{thebibliography}

\begin{thebibliography}{99}
\bibitem{linoSM}
Digregorio P., Levis D., Suma A., Cugliandolo L.F., Gonnella G., Pagonabarraga I., Phys. Rev. Lett. \textbf{121}, 098003 (2018).
\bibitem{braySM}
Bray A.J., Adv. in Phys. \textbf{51}, 481 (2002).
\bibitem{stenSM}
Stenhammar J., Marenduzzo D., Allen R.J., Cates M.E., Soft Matter \textbf{10}, 1489 (2014).
\bibitem{furuSM}
Furukawa H., Phys. Rev. B \textbf{40}, 2341 (1989).
\bibitem{esterSM}
Ester M., Kriegel H.-P., Sander J., Xu X., in \textit{Proceedings of the Second International Conference on Knowledge
Discovery and Data Mining} (1996).
\end{thebibliography}

\end{document}